\documentclass[pra,aps,nobalancelastpage,twocolumn,superscriptaddress,longbibliography]{revtex4-1} 
\pdfoutput=1
\usepackage[utf8]{inputenc}
\usepackage{color,amsthm,amsmath,amsxtra,amsfonts,dsfont,graphicx,bm}
\usepackage{float} 

\newcommand{\be}{\begin{equation}}
\newcommand{\ee}{\end{equation}}
\newcommand{\bea}{\begin{eqnarray}}
\newcommand{\eea}{\end{eqnarray}}
\newcommand\RR{\mathbf{R}}
\newcommand\nn{\mathbf{n}}
\newcommand\ii{\mathbf{i}}
\newcommand\jj{\mathbf{j}}

\newcommand\mm{\mathbf{m}}
\newcommand\sss{\mathbf{s}}

\newcommand\kk{\mathbf{k}}
\newcommand\qq{\mathbf{q}}

\newcommand\rr{\mathbf{r}}
\newcommand\xx{\mathbf{x}}

\newcommand\hc{\text{H.c.}}

\newcommand{\st}[1]{_\text{#1}}

\newcommand{\bra}[1]{\left\langle #1\right|}
\newcommand{\ket}[1]{\left| #1\right\rangle}
\newcommand{\lla}[1]{\left\{ #1\right \} }

\newcommand{\braket}[2]{\langle #1|#2\rangle}
\newcommand{\bkev}[3]{\langle #1|#2| #3 \rangle}
\newcommand{\pa}[1]{\left( #1\right)}

\newcommand{\abs}[1]{\left| #1\right|}

\newcommand{\fl}[1]{\left\lfloor #1\right\rfloor}

\newcommand{\dg}[0]{^\dagger}
\newcommand{\vac}[0]{\ket{\text{vac}}}
\newcommand{\mean}[1]{\left\langle #1\right\rangle}

\usepackage{hyperref}

\definecolor{mygrey}{rgb}{0,0.45,0.45}
\definecolor{myblue}{rgb}{0.2,0.2,0.8}
\definecolor{myzard}{cmyk}{0,0,0.05,0}
\definecolor{mywhite}{rgb}{1,1,1}
\definecolor{mywhite}{rgb}{1,1,1}
\definecolor{myred}{rgb}{1,0.,0.3}

\newcommand{\norm}[1]{\left|#1\right|}
\newcommand{\co}[1]{\left[ #1\right]}

\hypersetup{
  pdfsubject={},
  pdfcreator={pdflatex},
  colorlinks,
  linkcolor=myblue,
  citecolor=myblue, 
  urlcolor=myblue
}

\begin{document}

\title{Engineering analog quantum chemistry Hamiltonians using cold atoms in optical lattices}
\date{\today}

\begin{abstract}

\end{abstract}

\author{Javier Arg\"uello-Luengo}
\email{javier.arguello@icfo.eu}
\affiliation{ICFO-Institut de Ci\`encies Fot\`oniques, The Barcelona Institute of Science and Technology,
08860 Castelldefels (Barcelona), Spain}

\author{Tao Shi}
\email{tshi@itp.ac.cn}
\affiliation{CAS Key Laboratory of Theoretical Physics, Institute of Theoretical Physics, Chinese Academy of Sciences, P.O. Box 2735, Beijing 100190, China}

\author{Alejandro Gonz\'alez-Tudela}
\email{a.gonzalez.tudela@csic.es}
\affiliation{Instituto de F\'isica Fundamental IFF-CSIC, Calle Serrano 113b, Madrid 28006, Spain}

\begin{abstract}

Using quantum systems to efficiently solve quantum chemistry problems is one of the long-sought applications of near-future quantum technologies. In a recent work~\cite{arguello2019analogue}, ultra-cold fermionic atoms have been proposed for this purpose by showing us how to simulate in an analog way the quantum chemistry Hamiltonian projected in a lattice basis set. Here, we continue exploring this path and go beyond these first results in several ways. First, we numerically benchmark the working conditions of the analog simulator, and find less demanding experimental setups where chemistry-like behaviour in three-dimensions can still be observed. We also provide a deeper understanding of the errors of the simulation appearing due to discretization and finite size effects and provide a way to mitigate them. Finally, we benchmark the simulator characterizing the behaviour of two-electron atoms (He) and molecules (HeH$^+$) beyond the example considered in the original work.

\end{abstract}

\maketitle

\section{Introduction}
\label{sec:intro}
Solving quantum chemistry problems, such as obtaining the electronic structure of complex molecules or understanding chemical reactions, is an extremely challenging task. Even if one considers the nuclei positions $\{\RR_\alpha \}_{\alpha=1}^{N_n}$ fixed due to their larger mass (Born-Oppenheimer approximation), and focus only on the electronic degrees of freedom, these problems still involve many electrons interacting through Coulomb forces, whose associated Hilbert space grows exponentially with the number of electrons ($N_e$). One way of circumventing this exponential \emph{explosion}~\cite{feynman82a} consists in using the electron density instead of the wavefunction, like in density-functional methods~\cite{hohenberg64a,kohn65a}, where the complexity is hidden in the choice of exchange-correlation density functionals. Educated guesses of such functionals have already allowed us to study the properties of large molecules~\cite{jones15a}. Unfortunately, there is no unambiguous path for improving these functionals~\cite{cohen08a,medvedev17a,kepp17a,medvedev17c,hammesschiffer17a,brorsen17a,korth17a,gould17a,mezei17a,wang17b,kepp18a,su18a}, which are known to fail in certain regimes~\cite{cohen08a}. A complementary route consists in projecting the quantum chemistry Hamiltonian in a basis set~\cite{szabo12a,lehtola19a} with a finite number of elements $N_o$. The typical choices for the basis are linear combinations of atomic orbitals with Slater- or gaussian-type radial components. These methods generally provide good accuracies with small $N_o$. However, the quality of the solution ultimately depends on the basis choice.
And on top of that, the Hilbert space of the projected Hamiltonian still grows exponentially with $N_o$, which complicates their solution if large basis sets are required, especially for non-equilibrium situations.

In parallel to these developments, the last few years have witnessed the emergence of an alternative route to study these problems based on using quantum systems to perform the computation. This idea was first proposed by Feynman as a way of preventing the exponential explosion of resources of quantum many-body problems~\cite{feynman82a}, formalized later by Lloyd~\cite{lloyd96a}, and first exported into the quantum chemistry realm by Aspuru-Guzik \emph{et al}~\cite{aspuruguzik05a}. First algorithms used Gaussian orbital sets and phase-estimation methods to obtain ground-state molecular energies~\cite{whitfield11a,wecker14a}. Despite the initial pessimistic scaling of the gate complexity with the number of orbitals (polynomial but with a large exponent), recent improvements through the use of more efficient algorithms~\cite{Berry2019a} or different basis sets, e.g., plane-waves~\cite{babbush18a,babbush2019quantum,Low2019a}, have reduced significantly the gate scaling complexity. Since these algorithms typically assume fault-tolerant quantum computers that will not be available in the near-future, in the last years there has also been an intense effort on hybrid variational approaches more suitable for current noisy quantum computers~\cite{peruzzo14a,omalley16a,kandala17a}. However, these will be ultimately limited by the available ans\"atze that can be obtained with current devices, as well as on the optimization procedure~\cite{Wang2020,Czarnik2020}.

The previously described efforts (see Ref.~\cite{Cao2019a} for an updated review) fall into what is called the digital quantum simulation framework, in which the fermionic problem is mapped into qubits and the Hamiltonian evolution is performed stroboscopically. In a recent work~\cite{arguello2019analogue}, the authors and co-workers opened a complementary route to study these problems showing how to simulate in an analog way the quantum chemistry Hamiltonian using a discretized space  (or grid) basis representation~\cite{white89a}. These representations have been generally less used in the literature due to the large basis sets required to obtain accurate results. However, they have recently experienced a renewed interest~\cite{white17a,stoudenmire17a,white19a} due to their better suitability for DMRG methods~\cite{white04a}. In our case~\cite{arguello2019analogue}, we use this grid representation because it is well suited for describing fermions trapped in optical lattice potentials, where the fermionic space is naturally \emph{discretized} in the different trapping minima of the potential. Then, as explained in Ref.~\cite{arguello2019analogue}, fermionic atoms with two internal atomic states can hop around the lattice, playing the role of  electrons, spatially-shaped laser beams simulate the nuclear attraction, while an additional auxiliary atomic specie mediates an effective repulsion between the fermionic atoms that mimics the Coulomb repulsion between electrons. In this work, we continue exploring the path opened by Ref.~\cite{arguello2019analogue} and extend its results. The manuscript is structured as follows:
\begin{itemize}
	\item In Section~\ref{sec:qcHamiltonian} we introduce the different parts of the quantum chemistry Hamiltonians projected in finite basis sets. We discuss both the grid basis representation that we use in our analog simulation and the widely-used linear combination of atomic orbitals, emphasizing the similarities and the differences between these two approaches.
	
	\item In Section~\ref{sec:qcsingle} we review how to obtain the single-particle parts of the quantum chemistry Hamiltonian, that are, the electron kinetic energy and the nuclear attraction, as proposed in Ref.~\cite{arguello2019analogue}. Besides, we extend the previous analysis with a deeper understanding of the discretization and finite size errors of the simulation, which allows us to introduce an extrapolation method that mitigates the limitations imposed by these errors for a given lattice size.
	
	\item In Section~\ref{sec:qcelectron} we analyze the role of the different ingredients introduced in Ref.~\cite{arguello2019analogue} to obtain an effective pair-wise and Coulomb-like repulsion between the fermionic atoms. This analysis enables us to present less demanding simplified experimental setups to simulate chemistry-like behaviour in three-dimensional systems. Besides, we numerically benchmark the parameter regimes where the analog simulator works beyond the perturbative analysis of Ref.~\cite{arguello2019analogue}.
	
	\item In Section~\ref{sec:benchmark}, we put all the ingredients together and benchmark our simulator beyond the example considered in Ref.~\cite{arguello2019analogue}, i.e., considering two-electron atoms (He) and molecules (He$^+$-H).
	
	\item Finally, in Section~\ref{sec:conclusion} we summarize our findings and point to further directions of work.
\end{itemize}

\section{Quantum chemistry Hamiltonians in discrete basis sets: atomic orbitals \emph{vs.} grid basis}
\label{sec:qcHamiltonian}

The typical problems in quantum chemistry are either calculating the electronic structure of a complex molecule in equilibrium, $\hat{H}_e\ket{\Psi}_e=E_e\ket{\Psi}_e$, or its time-evolution in an out-of-equilibrium situation: $i\partial_t\ket{\Psi(t)}_e=\hat{H}_e\ket{\Psi(t)}_e$. These problems are generally calculated using the Born-Oppenheimer approximation (BOA), that is, treating each nuclei classically as a fixed particle of charge $Z_\alpha e$. Along the manuscript, we will use atomic units $m_e=e=\hbar=(4\pi\varepsilon_0)^{-1}\equiv 1$, such that the natural unit of length will be given by the Bohr Radius $\text{a}_0=(4\pi\varepsilon_0 \hbar^2)/(m_e e^2)\equiv 1$, and the unit of energy is the Hartree-Energy $E_h=\hbar^2/(m_e \text{a}_0^2)\equiv 1$ (twice the Rydberg energy (Ry)). Using these units, the BOA-electronic Hamiltonian for a molecule with $N_e$ electrons and a given nuclei configuration $\{\RR_\alpha\}_{\alpha=1}^{N_n}$ reads:
\begin{subequations}
\begin{align}
\hat{H}_e=&\sum_{j=1}^{N_e}\left[-\frac{1}{2}\hat{\nabla}_j^2-\sum_{\alpha=1}^{N_n}Z_\alpha \hat{V}_c(\hat{\rr}_j,\RR_\alpha)\right]\label{subeq:Hsingle}\\
&+\frac{1}{2}\sum_{i\neq j=1}^{N_e} \hat{V}_c(\hat{\rr}_i,\hat{\rr}_j)=\label{subeq:Htwo}\\
&=\sum_{j=1}^{N_e}  \hat{H}_1(\hat{\rr}_j)+\frac{1}{2}\sum_{i\neq j=1}^{N_e} \hat{V}_c(\hat{\rr}_i,\hat{\rr}_j)\,,
\end{align}
\end{subequations}
where bold letters indicate three-dimensional vectors and $V_c(\mathbf{r}_1,\mathbf{r}_2)=\frac{\text{a}}{|\mathbf{r}_1-\mathbf{r}_2|}$ is the pair-wise Coulomb potential between the charged particles (electrons and nuclei). The expression inside the brackets of Eq.~\eqref{subeq:Hsingle} is labeled as the single-electron part of the Hamiltonian ($\hat{H}_1(\hat{\rr}_j)$)  and contains both the electron kinetic energy ($\hat{T}_e=-\sum_{j}\nabla_j^2/2$) and the electron-nuclei attraction ($\hat{H}_{n-e}=-\sum_{\alpha=1}^{N_n}Z_\alpha \hat V_c(\hat{\rr}_j,\RR_\alpha))$, whereas Eq.~\eqref{subeq:Htwo} corresponds to the electron-electron repulsion ($\hat{V}_{e-e}$).

Since the molecular electrons are indistinguishable (up to the spin degree of freedom), for computational purposes it is typically more convenient to write a second-quantized version of the Hamiltonian that already takes into account the fermionic statistics of the particle. There is a general recipe to do it~\cite{mahanbook13a,galindo2012quantum}: first, one needs a set of single-particle states $\mathcal{B}=\{\ket{\phi_i}\}$, that can be used to define an abstract Hilbert space of states $\ket{n_1,n_2,\dots}$, denoting that there are $n_i$ electrons occupying the $i$-th single-particle states. With these states, one can then define annihilation (creation) operators $\hat{c}_{i}^{(\dagger)}$ that denote the creation/destruction of a fermionic particle in the $i$-th single-particle state. This labelling already accounts for the different spin states and the fermionic statistics of the particle through their anticommutation rules: $\{\hat{c}_{i},\hat{c}_{j}^\dagger\}=\delta_{ij}$, and $\{\hat{c}_{i},\hat{c}_{j}\}=\{\hat{c}^\dagger_{i},\hat{c}_{j}^\dagger\}=0$. With these operators, one can define the field operators: 
\begin{subequations}
\begin{align}
    \label{eq:field}
\hat{\Psi}(\rr)&=\sum_{i} \phi_i(\rr) \hat{c}_{i}\,,\\
\hat{\Psi}^{\dagger}(\rr)&=\sum_{i} \phi^{*}_i(\rr) \hat{c}_{i}^{\dagger}\,,
\end{align}
\end{subequations}
that can be used to write the Hamiltonian in the following form:
\begin{subequations}
\begin{align}
\hat{H}_e&=\int d\rr \hat{\Psi}^\dagger(\rr)\left[-\frac{1}{2}\hat{\nabla}^2-\sum_{\alpha=1}^{N_n} \frac{Z_\alpha}{|\rr-\RR_\alpha|}\right]\hat{\Psi}(\rr)  \label{eq:Hesecond1}\\
&+\frac{1}{2}\iint d\rr d\rr'\hat{\Psi}^\dagger(\rr)\hat{\Psi}^\dagger(\rr')\frac{1}{|\rr-\rr'|} \hat{\Psi}(\rr')\hat{\Psi}(\rr)\label{eq:Hesecond2}\,.
\end{align}
\end{subequations}
If the basis $\mathcal{B}$ of single-particle states is complete (i.e., it is infinite dimensional), the mapping between the first quantized Hamiltonian of Eqs.~\eqref{subeq:Hsingle}-\eqref{subeq:Htwo} and the second quantized one of Eqs.~\eqref{eq:Hesecond1}-\eqref{eq:Hesecond2} would be exact. However, this is generally not practical since the associated Hilbert space will still be infinite. For those reasons, the typical approach consists in projecting the Hamiltonian in the subspace spanned by the tensor product of a finite-dimensional discrete basis set, $\mathcal{B}_t$, and solving the problem within that subspace. The prototypical bases chosen are built out of (linear combinations) of atomic orbitals centered around the nuclei position, labeled as linear combination of atomic orbitals (LCAO) basis sets~\cite{szabo12a}. However, for our analog quantum chemistry simulation it will be more adequate to use an alternative representation based on a grid discretization of the continuum in a finite set of points. In what follows, we discuss how the second quantized electronic Hamiltonian looks in both cases, and highlight their main differences.

\emph{Linear combination of atomic orbitals (LCAO).} Here, the basis set is composed of $N_o (>N_e)$ single-particle (orthonormal) atomic orbitals, $\mathcal{B}_t=\{\ket{\phi_i}\}_{i=1}^{N_o}$, with which the Hamiltonian reads:
\begin{align}
\label{eq:Hesec}
\hat{H}_e=\sum_{i,j=1}^{N_o} t_{ij}\hat{c}^\dagger_{i}\hat{c}_{j}+\sum_{i,j.k,l=1}^{N_o} \frac{V_{ijkl}}{2} \hat{c}^\dagger_{i}\hat{c}^\dagger_{j}\hat{c}_{l}\hat{c}_{k}\,,
\end{align}
where the parameters of the discrete Hamiltonian $t_{ij}$ and $V_{ijkl}$ can be computed using the real space representation of the orbitals, $\phi_i(\rr)=\braket{\rr}{\phi_i}$, as follows:
\begin{subequations}
\begin{align}
\label{eq:discretet}
t_{ij}&=\int d\rr \phi^*_i(\rr)\left[-\frac{\nabla^2}{2}-\sum_{\alpha}Z_\alpha V_c(\mathbf{r},\mathbf{R}_\alpha)\right]\phi_j(\rr)\,,\\
V_{ijkl}&=\iint d\rr d\rr'\phi^*_i(\rr) \phi^*_j(\rr')\frac{1}{|\rr-\rr'|}\phi_l(\rr') \phi_k(\rr)\,,
\end{align}
\end{subequations}

The number of $t_{ij}$- and $V_{ijkl}$-parameters scales with the size of the $\mathcal{B}_t$-basis as $N_o^2$ and $N_o^4$, respectively, while their value depends on the particular states chosen. Convenient choices widely-used in quantum chemistry are linear combinations of Gaussian or exponential type-orbitals localized around the nuclei~\cite{szabo12a,lehtola19a}. The former are particularly appealing since the properties of Gaussian functions can simplify substantially the calculations of $t_{ij}, V_{ijkl}$, which can become a bottleneck if large basis sets are required. 

An advantage of this approach is that the number of orbitals required typically scales proportionally with $N_e$. Besides, it is a variational method that provides an unambiguous path to reach to the true ground state energy by increasing $N_o$. This is why these representations have been the most popular ones in most current approaches for digital quantum simulation~\cite{Cao2019a}. On the down side, the accuracy of the solution will depend on the particular molecular structure since the basis sets are composed of functions with fixed asymptotic decays that might not be suitable, e.g., to describe diffuse molecules~\cite{woon94a,sim92a,helgaker98a,jensen08a,lehtola12a,lehtola13a}.

\emph{Local or grid-discretized basis.} This option consists in writing the continuum Hamiltonian $\hat{H}_e$ in grid points $\nn=(n_x,n_y,n_z)\text{a}$, where $\text{a}$ is the spacing between the discretized points, $n_i\in \mathbb{Z}$, and $N=N_x N_y N_z$ being the total number of points. To do it, one can approximate the derivatives of the kinetic energy term in Eq.~\ref{subeq:Hsingle} by finite-differences, and evaluate the potentials at the grid points. This ultimately results in a second quantized Hamiltonian with the following shape~\cite{white89a,lehtola19a}:
\begin{subequations}
\begin{align}
\hat{H}_e&=-\sum_{\nn,\mm,\sigma}J_{\nn,\mm}\hat{c}^\dagger_{\nn,\sigma}\hat{c}_{\mm,\sigma}\label{subeq:kin}\\
&-\sum_{\alpha=1}^{N_n}\sum_{\nn,\sigma}{V}_\nn(\RR_{\alpha})\hat{c}^\dagger_{\nn,\sigma}\hat{c}_{\nn,\sigma}\label{subeq:nuc}\\
&+\frac{1}{2}\sum_{\nn, \mm,\sigma, \sigma'}{V}_\mathrm{el}(\nn,\mm) \hat{c}^\dagger_{\nn,\sigma}\hat{c}^\dagger_{\mm,\sigma'}\hat{c}_{\nn,\sigma}\hat{c}_{\mm,\sigma'}\label{subeq:rep} \,,
\end{align}
\end{subequations}
where $\hat{c}^{(\dagger)}_{\nn,\sigma}$ are now the local operators creating an electron with spin $\sigma$ at site position $\nn$, satisfying $\{\hat{c}_{\mm,\sigma},\hat{c}_{\nn,\sigma'}^\dagger\}=\delta_{\mm\nn}\delta_{\sigma,\sigma'}$. The kinetic energy coefficients [in Eq.~\eqref{subeq:kin}] $J_{\nn,\mm}$ depend on the expansion order chosen to approximate the Laplacian, and decay with the separation between sites $|\nn-\mm|$. For this manuscript, we will use the simplest finite difference formula for the second order derivative: 
\begin{align}
\label{eq:finite}
	\frac{d^2 f(x)}{d x^2}\approx \frac{f(x+\text{a})-2 f(x)+f(x-\text{a})}{\text{a}^2}\,,
\end{align}
which means that only nearest neighbour hoppings (and on-site energy) will appear in the kinetic energy term of Eq.~\eqref{subeq:kin}, and $J_{\nn,\mm}\equiv 0$ for the rest of the hopping terms. The nuclei-attraction term [Eq.~\eqref{subeq:nuc}] induces a position-dependent energy shift on the discretized electron orbitals coming from the attraction of the nuclei. Finally, the electron-electron repulsion [Eq.~\eqref{subeq:rep}] translates into long-range density-density interactions between the localized fermionic states. In the limit where $N\rightarrow \infty$ and $\text{a}\rightarrow 0$, the Hamiltonian of Eqs.~\eqref{subeq:kin}-\eqref{subeq:rep} converges to the continuum one.

This method typically requires larger basis sets to obtain accurate results~\cite{white89a} compared to LCAO ones. However, the number of interaction terms ${V}_\mathrm{el}(\nn,\mm)$ scales quadratically with the size of the basis because only density-density interaction terms appear. This can yield dramatic improvements when applying tensor-network methods, which motivates the renewed interest they have experienced in the last years~\cite{white17a,stoudenmire17a,white19a}. Besides, for the analog quantum simulation perspective such density-density interactions appear more naturally than the four-index interactions appearing in LCAO approaches.

A potential disadvantage is that these methods are generally not variational. That is, increasing $N_o$ might sometimes yield a larger energy than the one of smaller basis sets. This has been identified as a problem of underestimation of the kinetic energy when using the finite-difference approximation of the derivates [Eq.~\eqref{eq:finite}]~\cite{maragakis01a}. However, there are constructive ways of making the kinetic operator variational using different approximations of the kinetic energy~\cite{maragakis01a}. Along this manuscript, however, we will stick to the simple finite-difference formula of Eq.~\eqref{eq:finite} because of its simplicity. Besides, we will also provide a way of mitigating such discretization errors using an extrapolation method that we discuss in section~\ref{subsec:errors}.

In what follows, we explain how to simulate the different parts of the quantum chemistry Hamiltonian projected in a grid basis using ultra-cold atoms trapped in optical lattices, as initially proposed in Ref.~\cite{arguello2019analogue}. The reason for choosing this platform is that fermionic atoms with (at least) two internal atomic states can be used to describe electrons without the need to encode these operators into qubits, simplifying the Hamiltonian simulation, as already pointed out in earlier proposals~\cite{Luhmann2015,Sala2017,Senaratne2018}. We start by considering the single-particle part of the Hamiltonian in Section~\ref{sec:qcsingle}, and then explain how to obtain the electron repulsion in Section~\ref{sec:qcelectron}.

\section{Simulating single-particle Hamiltonian with ultra-cold atoms in optical lattices: kinetic and nuclear energy terms}
\label{sec:qcsingle}

The dynamics of ultra-cold fermionic atoms trapped in optical lattices is described by the following first-quantized Hamiltonian:
\begin{align}
\label{eq:fermHamiltonian}
\hat{H}_{f}=\hat{T}_f+\hat{V}_{\mathrm{per}}(\hat{\rr})+\hat{V}_{\mathrm{aux}}(\hat{\rr})\,,
\end{align}
which contains three terms: 
\begin{itemize}
\item The kinetic energy of the fermionic atoms, which reads:
\begin{align}
\hat{T}_f=-\frac{\hbar^2}{2 M_f}\sum_{j=1}^{N_f}\hat{\nabla^2}_j\,,
\end{align}
where $N_f$ is number of fermions of the system (that should be equal to the number of electrons we want to simulate), and $M_f$ their mass.

\item The optical lattice potential $\hat{V}_{\mathrm{per}}(\rr)$. This is typically generated by retro-reflected laser beams with a wavelength $\lambda$ that is off-resonant with a given atomic optical transition. These lasers form standing-waves which generate a spatially periodic energy shift~\cite{Grimm2000}, whose amplitude can be controlled through the laser intensity and/or detuning from the optical transition. Assuming a cubic geometry for the lasers, the optical potential reads~\cite{bloch08a}:
\begin{align}
V_{\mathrm{per}}(\rr)=-V_D\sum_{\alpha=x,y,z} \sin^2\left(\frac{2\pi \alpha}{\lambda}\right)\,,\label{eq:potential}
\end{align}
where $V_D$ is the trapping potential depth, that we assume to be equal for the three directions. The lattice constant of such potential is $\text{a}=\lambda/2$, and it imposes a maximum kinetic energy of the fermions $E_R=\hbar^2\pi^2/(2 M_f \text{a}^2)$, typically labeled as \emph{recoil energy}, which is the natural energy scale of these systems. Note that because of the larger mass of the fermions compared to the real electron systems, the dynamics will occur at a much slower timescale (ms) compared to electronic systems (fs). This will facilitate the observation of real-time dynamics of the simulated chemical processes, something very difficult to do in real chemistry systems.

\item Finally, we also included $V_\mathrm{aux}(\rr)$ that takes into account all possible atomic potential contributions which are not periodic, as it will be the case of the nuclear attractive potential.

\end{itemize}

Note that when writing only these three contributions for the fermionic Hamiltonian $\hat{H}_f$, we are already assuming to be in the regime where inter-atomic interactions between the fermions are negligible. This regime can be obtained, e.g., tuning the scattering length using Feshbach resonances~\cite{chin10a}. 

As for the quantum chemistry case, here it is also convenient to write a second-quantized version of the Hamiltonian $\hat{H}_f$. For that, we need to first find an appropriate set of single-particle states to define the field operators $\hat{\Psi}_f(\rr)$ as in Eq.~\eqref{eq:field}, and afterward write the second-quantized Hamiltonian using them.
In what follows, we explain the steps and approximations in the canonical approach to do it, whose details can be found in many authoritative references~\cite{jaksch98a,bloch08a,esslinger10a}.  

First, it is useful to characterize the band-structure emerging in the single-particle sector due to the potential $\hat{V}_\mathrm{per}(\rr)$. Since $\hat{V}_\mathrm{per}(\rr+\RR)=\hat{V}_\mathrm{per}(\rr)$ for $\RR=\sum_{i=1}^3 n_i \mathbf{a}_i$, with $\mathbf{a}_{1,2,3}=\mathrm{a}\hat{x},\mathrm{a}\hat{y},\mathrm{a}\hat{z}$ and $n_i\in \mathbb{Z}$, we can use the Bloch theorem to write the single-particle eigenstates of $\hat{H}_f$ as follows:
\begin{equation}
\psi_{n,\qq}(\rr)=u_{n,\qq}(\rr) e^{i\qq\cdot\rr}\,,\label{eq:bloch}
\end{equation}
where $\qq$ is the quasimomentum in the reciprocal space, $u_{n,\qq}(\rr)$ is a function with the same periodicity than $V_\mathrm{per}(\rr)$, and $n$ is denoting the index of the energy band $E_{n}(\qq)$. In the limit where the trapping potential depth is much larger than the recoil energy ($V_D/E_R\gg 1$), the atomic wavefunctions become localized in the potential minima. This is why, in that limit, it is useful to adopt a description based on Wannier functions localized in each potential minima, instead of the Bloch states $\phi_{\qq,n}(\rr)$ of Eq.~\eqref{eq:bloch}. The Wannier function of a site $\jj$ for the $n$-th band can be obtained from $\phi_{n,\qq}(\rr)$ as follows:
\begin{align}
W_{n,\jj}(\rr)=\frac{1}{\sqrt{N}}\sum_{\qq\in\mathrm{BZ}}\phi_{n,\qq}(\rr)e^{-i\jj\qq} \,,
\end{align}
where $N$ is the total number of sites of the optical potential.
In the strong-confinement limit,  $V_D\gg E_R$, the atoms only probe the positions close to the minima where the periodic potential can be expanded to $V_\mathrm{per}(\rr)=V_D \pi^2 r^2/\text{a}^2$, where $r=|\rr|$. This allows one to obtain an analytical expression for the Wannier functions in this limit in terms of the eigenstates of an harmonic potential with trapping frequency:
\begin{align}
\omega_{t,\alpha}=\frac{\sqrt{4 V_D E_R}}{\hbar}\,,
\end{align}
with $\alpha=x,y,z$, which provides an energy estimate for the energy separation between the different bands appearing in the structure.  Now, to write the field operator $\hat{\Psi}_f(\rr)$, one typically assumes that the atoms are prepared in the motional ground state of each trapping minimum, and that interband transitions are negligible~\cite{jaksch98a,bloch08a,esslinger10a}. With these assumptions, $\hat{\Psi}_f(\rr)$ can be expanded only in terms of states of the lowest energy band:
\begin{align}
\hat{\Psi}_f(\rr)=\sum_{\jj,\sigma} W_{\jj}(\rr) \hat{f}_{\jj,\sigma}\,,
\end{align}
where we drop the band-index $n$, and where we define the annihilation (creation) operators $\hat{f}^{(\dagger)}_{\jj,\sigma}$ of a fermionic state with spin $\sigma$ at site $\jj\in \mathbb{Z}^3$~\footnote{We assume that the lattice and nuclei positions are normalized to the lattice constant $d$, such that $V_0$ has units of energy.}, which also obey anti-commutation rules $\{\hat{f}_{\ii,\sigma},\hat{f}^\dagger_{\jj,\sigma'}\}=\delta_{\ii,\jj}\delta_{\sigma,\sigma'}$. With these operators, the second quantized fermionic Hamiltonian, $\hat{H}_f$, reads:
\begin{equation}
\hat{H}_f=-\sum_{\ii,\jj,\sigma} t_{\ii,\jj}\hat{f}_{\ii,\sigma}^\dagger \hat{f}_{\jj,\sigma}+\sum_{\ii,\sigma} \varepsilon_\jj \hat{f}_{\ii,\sigma}^\dagger \hat{f}_{\jj,\sigma}\,,\label{eq:Hfsec}
\end{equation}
where $t_{\ii,\jj}$ is the tunneling between the sites $\{\ii,\jj\}$ induced by the kinetic energy of the atoms, and $\varepsilon_\jj$ a position dependent energy shift coming from $\hat{V}_\mathrm{aux}(\rr)$. Note that these terms resemble the single-particle part of the quantum chemistry Hamiltonian $\hat{H}_e$ of Eqs.~\eqref{subeq:kin}-\eqref{subeq:nuc}. In what follows, we analyze in detail both terms, and explain how to make them match exactly those of Eqs.~\eqref{subeq:kin}-\eqref{subeq:nuc}.

\subsection{Electron kinetic energy}
\label{subsec:kinetic}
Equivalently to Eq.~\eqref{eq:discretet}, the strength of tunneling amplitude matrix $t_{\ii,\jj}$ of Eq.~\eqref{eq:Hfsec} is given by~\footnote{The on-site tunneling term $t_{\ii,\ii}=\hbar\omega_t$. Since this is a constant energy term in all lattice sites this is typically taken as the energy reference and set to $0$}:
\begin{align}
t_{\ii,\jj}=\int d\rr W_\ii^*(\rr)\left[-\frac{\hbar^2}{2 M_f}\nabla^2+V_\mathrm{per}(\rr)\right]W_\jj(\rr)\,.
\end{align}
In the strong-confinement limit that we are interested in, the Wannier functions are strongly localized around the minima (as Gaussians), such that in practical terms only nearest neighbour contributions appear. The strength of the nearest neighbour hopping terms can be estimated within this limit calculating the overlap between the Wannier functions as~\cite{bloch08a}:
\begin{align}
t_{\mean{\ii,\jj}}\equiv t_f\approx  E_R\sqrt{\frac{4}{\pi}}\left(\frac{V_D}{E_R}\right)^{3/4} e^{-2\sqrt{V_D/E_R}}\,,
\end{align}
where $\mean{\ii,\jj}$ denotes nearest-neighbor positions in the lattice. 
The kinetic part of the ultra-cold fermionic atoms in optical lattices is then approximated by:
\begin{align}
\hat{T}_f=-t_f\sum_{\mean{\ii,\jj},\sigma} \hat{f}_{\ii,\sigma}^\dagger \hat{f}_{\jj,\sigma}\,.
\end{align}
This gives exactly the electron kinetic energy terms of Eq.~\eqref{subeq:kin} using the finite-difference approximation of the derivative of Eq.~\eqref{eq:finite}, up to a constant energy shift $2t_f$ that commutes with the complete Hamiltonian.

\subsection{Nuclear attraction}
\label{subsec:nuclear}

\begin{figure}[tbp]
	\centering
	\includegraphics[width=0.9\linewidth]{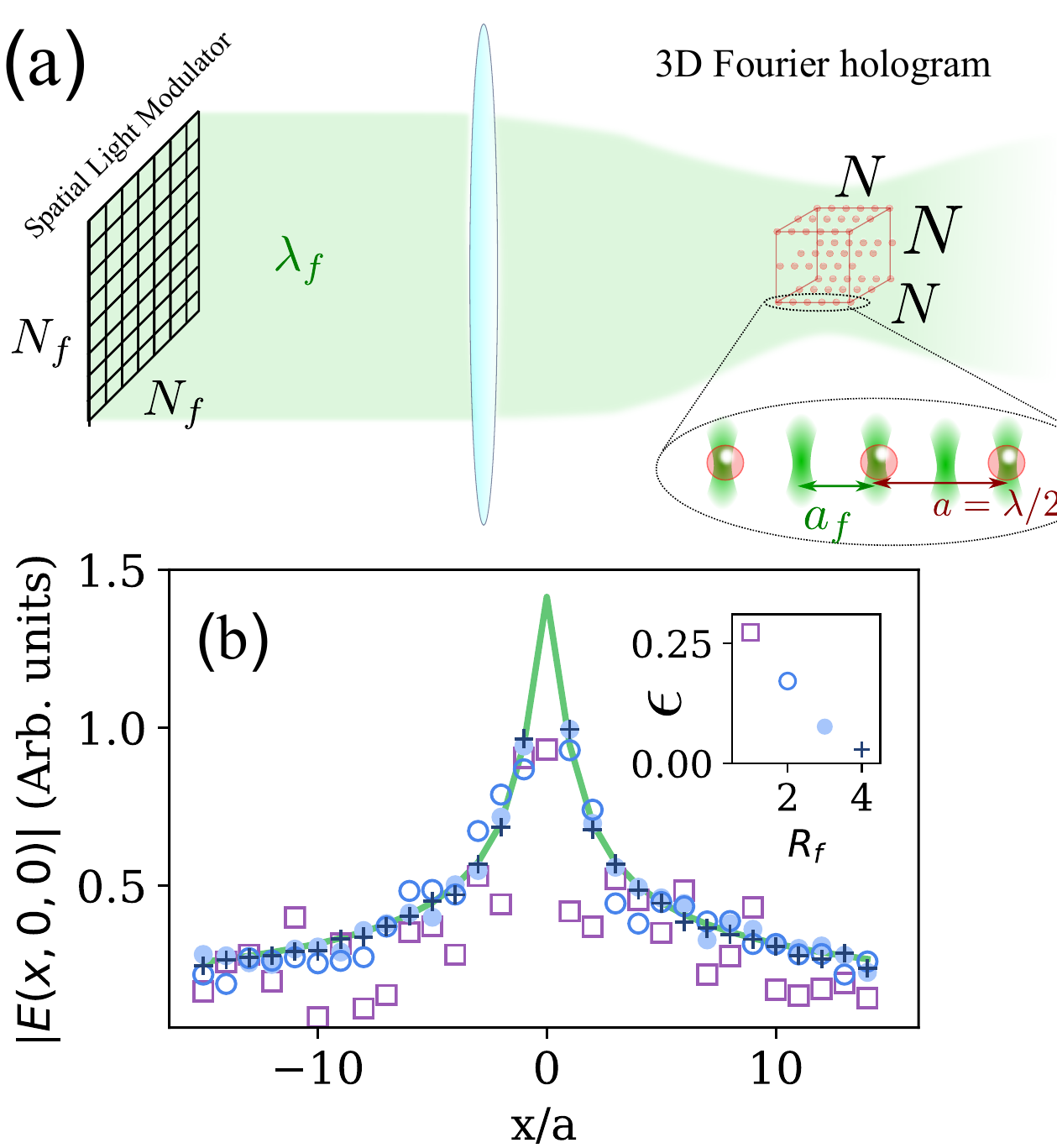}
	\caption{(a) Schematic representation of the holographic approach. A monochromatic laser beam with wavelength $\lambda_f$ is reflected into a spatial light-modulator (SLM) which imprints a complex phase pattern in a two-dimensional grid with $N_f\times N_f$ sampling points. A lens creates a 3D Fourier hologram in a volume around its focal point where the fermionic optical lattice of $N\times N\times N$ sites and spacing $\text{a}$ is placed. The minimum distance in which the electric field can be modulated in the 3D hologram, $\text{a}_f$, depends on the optical properties of the setup, but it is always lower-bounded by the diffraction limit of light $\text{a}_f\geq \lambda_f/2$. As shown in Ref.~\cite{whyte05a}, intensity of the field at each point is determined by the Fourier Transform of the phase mask in the corresponding direction. (b) Linear cut of one of the calculated electric field amplitude at the center of the lattice after the iteration process of the G-S algorithm for refining factors $R_f\in(1-4)$ (see text for definition). Green line follows the desired Coulomb potential, $V_0/|\rr|$, and different markers are used for each refining factor, as indicated in the inset. \emph{Inset} shows the average normalized error for 30 random initializations of the algorithm. }
	\label{fig:holo_bigger}
\end{figure}

The nuclear attraction term of Eq.~\eqref{subeq:nuc} can be simulated by the position dependent shift $\varepsilon_{\jj}$, whose expression in terms of the Wannier functions reads as:
\begin{align}
\varepsilon_\jj=\int d\rr |W_\jj(\rr)|^2 V_\mathrm{aux}(\rr)\approx V_\mathrm{aux}(\jj) \,.
\end{align}

Thus, in order to match the nuclear attraction term of the quantum chemistry Hamiltonian of Eq.~\eqref{subeq:nuc}, we \emph{just} require that $V_\mathrm{aux}(\rr)$ has the shape of the nuclear Coloumb attraction, at least, at the optical lattice minima $\jj$ where the fermions can hop. To obtain that, one can add a red-detuned spatially shaped electric field beam, $\mathbf{E}_\alpha(\rr)$, for each of the nuclei we want to simulate, such that the induced light-shift generates an optical potential with the shape:
\begin{align}
V_\mathrm{aux}(\jj)=-\sum_{\alpha=1}^{N_n}\frac{|\mathbf{E}_\alpha(\jj)|^2}{\delta_\alpha}\approx -\sum_{\alpha=1}^{N_n} \frac{Z_\alpha V_0}{|\jj-\RR_{\alpha}|}\,,
\end{align}
with $V_0$ being the overall energy scale of the potential controlled by the intensity of the laser and/or detuning $\delta_\alpha$. For consistency of our model, the maximum energy difference between the different sites, that is of the order $\Delta\varepsilon_\mathrm{max}=Z_\mathrm{max} V_0$, should be much smaller than the trapping depth of the overall potential $\Delta\varepsilon_\mathrm{max}\ll V_D$, such that the tunnelings $t_f$ are not affected by it, and also $\Delta\varepsilon_\mathrm{max}\ll \hbar\omega_t$ so that it does not create interband transitions (see~\cite{hooley04a,rigol04a,rey05a,block10a} where similar effects were considered due to the existence of confinement potentials). Both limits can be satisfied in the regime of parameters we are interested in.

The non-trivial part here consists in obtaining the electric fields $\mathbf{E}_\alpha(\rr)$ with the desired intensity pattern, $|\mathbf{E}_\alpha(\rr)|^2$. One option could be harnessing the advances in 3D holographic techniques that allow to shape the electromagnetic field in a given volume by imprinting complex phase patterns in a two-dimensional grid and using Fourier optics to propagate them to the position of the optical lattice~\cite{goodman2005introduction}. These 3D holograms have already enabled, for example, trapping Rydberg atoms in exotic three-dimensional (3D) configurations~\cite{barredo18a}. The idea of holographic traps is conceptually simple (see Fig.~\ref{fig:holo_bigger}a): one impinges a monochromatic laser beam with wavelength $\lambda_f$ into a spatial light-modulator (SLM) that imprints a non-uniform phase pattern in a grid with $N_f\times N_f$ pixels. The reflected laser field is then focused with a high-numerical aperture lens to generate the desired 3D holographic intensity pattern~\cite{goodman2005introduction} that depends on the imprinted phases. The minimum spatial resolution ($\text{a}_f$) in which the electric field can be modulated depends on both the optical setup and the wavelength of the incident laser $\lambda_f$, but it will be always lower bounded by the diffraction limit of light $\text{a}_f\geq \lambda_f/2$. This motivates the use of high numerical aperture lenses~\cite{Bruno:19,Glicenstein2021,Beguin2020} to reduce the waist of the holographic beam. We will label as $R_f=\text{a}/\text{a}_f$ to the ratio between the inter-atomic distance in the optical lattice and the spatial resolution of the hologram.

The first step to design the 3D holograms consists in finding the appropriate phase pattern that should be imprinted in the $N_f\times N_f$ grid of the SLM to obtain the desired electric field intensity. Fortunately, there are many constructive algorithms of doing it~\cite{Pasienski2008, Wu2015,Chang2015}. Inspired by the original Gerchberg-Saxton (G-S) algorithm~\cite{gerchberg71a,shabtay03a,dileonardo07a,whyte05a}, here we follow the one of Ref.~\cite{whyte05a} adapted for modulating 3D electric fields in discrete points of space. This algorithm initially starts by a random set of phases in the $N_f\times N_f$, and then iteratively looks for a solution that both approximates a given intensity pattern at the fermionic positions, $V_\jj^0$, with the only restriction of satisfying Maxwell's equation. That is, the $\kk-$components of the beam of monochromatic light $\lambda_f$ lie in the Ewald sphere of radius $k_f=2\pi/\lambda_f$. The convergence of the solution can be monitored using the dimensionless factor: 
\begin{equation}
\label{eq:err}
\epsilon = \sum_\jj \abs{\frac{V_\jj-V^0_\jj}{V^0_\jj}}\,,
\end{equation}
where $V_\jj$ is the electric field intensity in position $\jj$ obtained at an iteration of the algorithm and $V_\jj^0$ the targetted one. A key element for the convergence of the algorithm is the number of sampling points $N_f$ of the grid, which for simplicity we will assume to be proportional to the number of optical lattice positions $N_f=R_f N$, choosing $R_f$ as the proportionality factor. Like this, if $R_f>1$ the hologram can find solutions where the electric field intensity is modulated also within the fermionic positions, which will facilitate the convergence of the algorithm.

In Fig.~\ref{fig:holo_bigger}(b) we plot the result of applying this algorithm~\cite{whyte05a} for several (integer) $R_f$ to the case of having a single nucleus at the origin position, that is, when $V_\jj^0$ should have a Coulomb shape potential around the origin~\footnote{We always consider that the nuclei are centered in a position separated half-a-lattice constant away from the lattice sites to avoid the divergent behaviour. This will introduce an error in our simulation, as we will consider more in detail in the next subsection.}. We apply the G-S algorithm to find the phase mask for a given (integer) $R_f$ until the improvement of the relative error $\epsilon$ from one iteration to the next is below $10^{-4}$. Then, we plot in the main panel a linear cut of the 3D electric field amplitude at the final iteration $V_\jj$ (with markers) compared to the targeted one $V^0_\jj$ (in solid green line), and its corresponding relative error $\epsilon$ in the inset. In purple squares we plot the case $R_f=1$ where we observe that the agreement with the desired potential leads to a final error above 10\%. However, as we increase the number of sampling points $N_f$ using larger $R_f$, the algorithm is able to find better solutions, as clearly indicated by the decrease of the final relative error as $R_f$ increases (see inset of Fig.~\ref{fig:holo_bigger}b). For example, with $R_f=4$ (blue crosses), the potential finally obtained captures very well the desired intensity profile at the positions of the fermions, obtaining a normalized relative error of $\epsilon\approx 0.02$. The most obvious way to increase $R_f$ consists in either increasing the optical lattice period or using smaller wavelengths for the focused laser. One option is to use Alkaline-Earth atoms which have a level structure that combines telecom transitions ($1.4$ $\mu$m) with ultra-violet ($400$ nm) ones~\cite{covey2019telecom}, although we recognize that going to large $R_f$ will be experimentally challenging and will require the use of innovative ideas, e.g., developing novel tweezers techniques~\cite{Beguin2020}. For this reason, in Section~\ref{subsec:errors} we will discuss the impact of imperfect potential in the precision of the simulators, showing how already $\varepsilon \sim 0.1$ can provide energy errors smaller than 1$\%$ for the simplest case of atomic hydrogen. 

\subsection{Errors: discretization, finite-size, and mitigation strategies}
\label{subsec:errors}

\begin{figure}[tbp]
	\centering
	\includegraphics[width=0.8\linewidth]{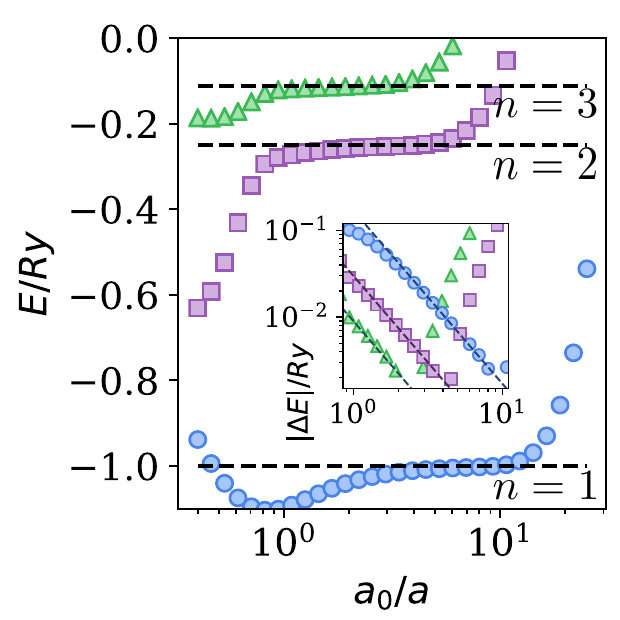}
	\caption{The lowest-energy states of the continuum (dashed black lines) and the discrete (color markers) Hydrogen atom associated to the principle quantum number $n=1,2,3$. Their energy depends on the effective Bohr radius. Once atomic orbitals occupy more than a single site, $\text{a}_0/\text{a}>1$, the spectrum approaches the analytical result $E=\text{Ry}/n^2$ (dashed lines), up to a critical size for which the finiteness of the lattice  becomes relevant. For $n=1$, these finite-size effects appear at around $a_0/a \approx 10$ for the present size. \emph{Inset}, shows the difference between the calculated and analytic energy associated to these curves. Dashed line follows the scaling expected from discretization effects, shown in Eq. \eqref{eq:discrete}. Lattice size, $N=100^3$.}
	\label{fig:hydrQual}
\end{figure}

Up to now, we have shown that the dynamics of ultra-cold fermionic atoms in deep optical lattices ($V_D\gg E_R$), and with an appropriate shaping of $V_\mathrm{aux}(\rr)$ can mimic the single-particle part of the quantum chemistry Hamiltonian:
\begin{align}
\hat{H}_f=-t_f \sum_{\mean{\ii,\jj},\sigma} \hat{f}^\dagger_{\ii,\sigma}\hat{f}_{\jj,\sigma}-\sum_{\alpha=1}^{N_n}\sum_{\jj,\sigma}\frac{Z_\alpha V_0}{|\jj-\RR_\alpha|}\hat{f}^\dagger_{\jj,\sigma}\hat{f}_{\jj,\sigma}\,,\label{eq:Hf3}
\end{align}

Before showing how to simulate the electron repulsion part of the quantum chemistry Hamiltonian (Eq.~\ref{subeq:rep}), in this section we will provide intuition on how the chemistry energies and length scales translate into the cold-atom simulation, and which are the errors appearing due to two competing mechanisms: discretization and finite size effects, taking as a case of study the Hydrogen atom. The reason for choosing this case is two fold: first, it can be simulated directly using the Hamiltonian of Eq.~\eqref{eq:Hf3} imposing $N_n=1$ and $Z_1=1$, since one does not require the electron interactions; second, it is fully understood analytically in the continuum limit, that will allow us to easily benchmark our results and define the natural units of our system. Identifying the discretized and continuum Hamiltonians, one can obtain the following correspondence:
\begin{align}
 Ry&=\frac{V_0^2}{4t_f}\,, \\
 \frac{\text{a}_0}{\text{a}}&=\frac{2 t_f}{V_0}\,,\label{eq:Bohr}
\end{align} 
where $Ry$ is the Rydberg energy of the simulated Hydrogen, and $\text{a}_0/\text{a}$ is the effective Bohr radius in units of the lattice constant. Since one can control the ratio $t_f/V_0$ at will with the lasers creating the optical potentials, one can effectively choose the Bohr radius of the discrete Hydrogen atom and, consequently, of the simulated molecules when more nuclei are present. This will be an important asset of our simulation toolbox since it will allow one to minimize the errors coming from discretization and finite-size effects. In order to illustrate it, we plot in Fig.~\ref{fig:hydrQual} the lowest energy spectrum of the discrete Hydrogen as a function of the effective Bohr radius $\text{a}_0/\text{a}$ defined in Eq.~\eqref{eq:Bohr}. The black dashed lines are the expected energies in the continuum Hamiltonian, i.e., $E_n=Ry/n^2$, whereas in the different colors are the different numerical energies for a fixed system size of $N=100^3$ sites.
From this calculation, we observe several features of the grid discretized basis that we are choosing to represent the quantum chemistry Hamiltonian. For example, when $\text{a}_0/\text{a}\lesssim 1$, all the states deviate from the expected energy. This is not surprising because in this regime, all the fermionic density is expected to concentrate around one trapping minimum, such that discretization effects become large. In the opposite regime, when the Bohr radius becomes comparable with system size, $\text{a}_0/\text{a}\gtrsim N^{1/3}$, the energies also deviate from the continuum result, since the discrete Hydrogen atom does not fit in our system.  Only in the intermediate regime one can minimize both errors and approximate well the correct energy. Note, however, that the optimal range of $\text{a}_0/\text{a}$ depends on the particular orbital considered. For example, the ground state $s$ orbital ($n=1$) is more sensitive to discretization effects since it has a larger fraction of atomic density close to the nucleus, while larger orbitals are more sensitive to finite size effects because their spatial extension grows with $n$.

This dependence of the convergence to the continuum limit on the particular atomic and molecular orbital will be commonplace in this method, and it also occurs for other basis representations~\cite{lehtola19a}. In spite of this, by analyzing the sources of errors one can extract some general conclusions that can provide valuable information when performing the experiments:
\begin{itemize}
	\item \emph{Discretization effects.} Analyzing numerically the convergence to the correct result in Rydberg units: $\Delta E=|E-E_\infty|$ (see inset in Fig.~\ref{fig:hydrQual}), we found an heuristic scaling of the error given by:
	\begin{align}
	\frac{\Delta E_{\mathrm{dis}}}{Ry}\propto \left(\frac{\text{a}}{\text{a}_0}\right)^2\propto \left(\frac{V_0}{t_f}\right)^2\,,
	\label{eq:discrete}
	\end{align}
	where the proportionality factor depends on the particular orbital studied. This scaling can be justified by considering the errors introduced by the discretization of the derivative and the integrals in the kinetic and potential energy term, respectively, leading both to the same scaling presented in Eq.~\eqref{eq:discrete} (see Appendix \ref{ap:discEffects}).
	
	\item \emph{Finite size effects.} These errors can be associated to the part of the electron density that can not be fitted within our system size. Since the Hydrogen orbitals decay exponentially with the principal quantum number $e^{-r /(n\text{a}_0)}$, one can estimate the errors due to finite size effect become exponentially smaller with the ratio between the system size and the orbital size, i.e.., $\propto e^{- N^{1/3}\text{a}/(n\text{a}_0)}$. 
\end{itemize}

\begin{figure}[tbp]
	\centering
	\includegraphics[width=\linewidth]{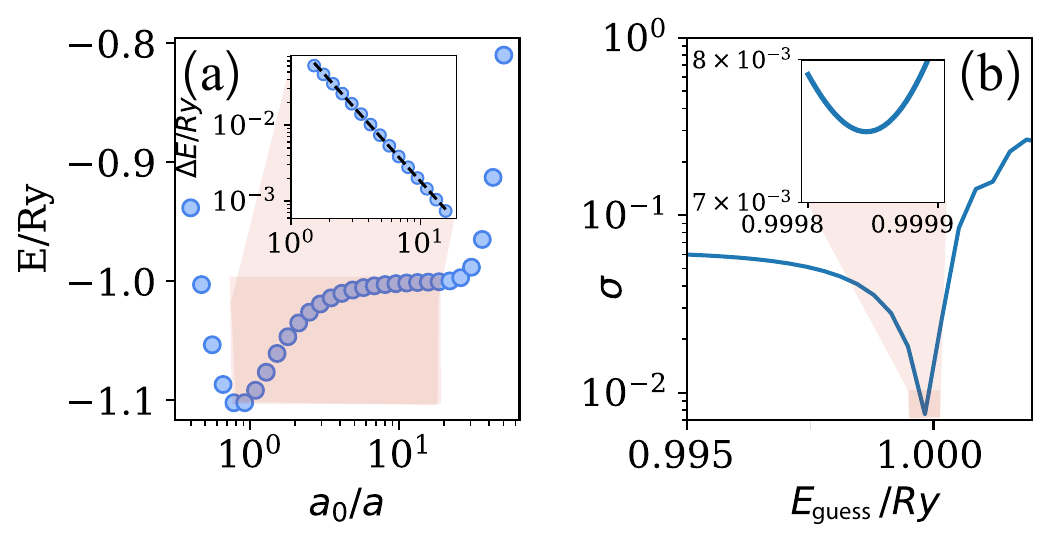}
	\caption{(a) Ground state of the discretized Atomic Hamiltonian depicted in Fig.~\ref{fig:hydrQual}, for different values of the effective Bohr radius and $N=250^3$. Inset: example of the fitting of the energy difference to the continuum value $\Delta E=E-E_\infty$, following a polynomial regression of the form $\Delta E/Ry=c \cdot (\text{a}_0/\text{a})^b$ for fitting parameters, $c$ and $b$. In this example we have chosen the continuum value $E_\infty$ to be $1Ry$. The best candidate for the value in the continuum needs to be calculated, as one generally does not know it a priori. (b) More systematically, here we show the standard deviation $\sigma$ in the determination of the fitting parameter $b$ for different candidate values of $E\st{guess}$.  We identify the best choice of $E\st{guess}/Ry$ as the one with the smallest deviation, $E_\infty$. In this simple example, one extrapolates the numerical result $E_\infty=0.9998 Ry$ (see inset), for the energy of $1s$ orbital. This is, a precision of $0.01\%$ for this simple scenario, gaining one order of magnitude with respect to the precision one can directly achieve before finite-size effects appear.}
	\label{fig:extrap}
\end{figure}

Even though these estimates were done based on numerical evidence of the Hydrogen atom, one can already extract important conclusions for the simulation of larger molecules. On the one hand, one can estimate the error scaling with electron density. Since each level with principal quantum number $n$ can fit $2 n^2$ electrons, an atom/molecule with $N_e$ electrons is expected to occupy a maximum quantum number $n_m\propto N_e^{1/3}$, such that its estimated size will be $L\propto N_e^{1/3} \text{a}_0/\text{a}$. Thus, following Eq.~\eqref{eq:discrete}, the discretization errors for such distances will scale with $\Delta E_\mathrm{dis}/Ry\propto \rho_e^{2/3}$, with $\rho_e=N_e/N$ the electron density. 

On the other hand, we can design an extrapolation method to obtain the energies with accuracies beyond the particular system size chosen and, importantly, without an \emph{a priori} knowledge of the exact result. We illustrate the method in Fig.~\ref{fig:extrap} for the ground state of Hydrogen, although in Section~\ref{sec:benchmark} we apply it as well to the case of multi-electron systems. The key steps go as follows: first, one calculates (or measures in the case of an experiment) the ground state energy for a fixed system size $N$ and for several ratios $\text{a}_0/\text{a}$ (panel a). Then, one defines $\Delta E/Ry=(E-E_\mathrm{guess})/Ry$ for several values of $E_\mathrm{guess}$ (panel b) and fit the resulting function to a polynomial regression $\Delta E/Ry=c (\text{a}_0/\text{a})^b$,  with free fitting parameters $b$ and $c$. We identify the right choice of the guess energy as the one with smallest standard deviation $\sigma$ (panel c), that we will say it is the one of the continuum limit $E_\infty\equiv E_\mathrm{guess}(\sigma_\mathrm{min})$. Using this procedure for a system size $N=250^3$, we obtain $E_\infty=0.9998Ry$, which is one order of magnitude better than the result one would obtain without extrapolation (i.e., directly looking at the minimum value of $E$ for that system size). For completeness, we also check that this value of $E_\infty$ also leads to an exponent factor compatible with $b=-2$ (not shown), which is the error scaling consistent with Eq.~\eqref{eq:discrete}. In Section~\ref{sec:benchmark}, this will be the criterion used to identify the best estimation for atomic and molecular energies beyond the discretization of the lattice. 

\begin{figure}[tbp]
\centering
\includegraphics[width=0.8\linewidth]{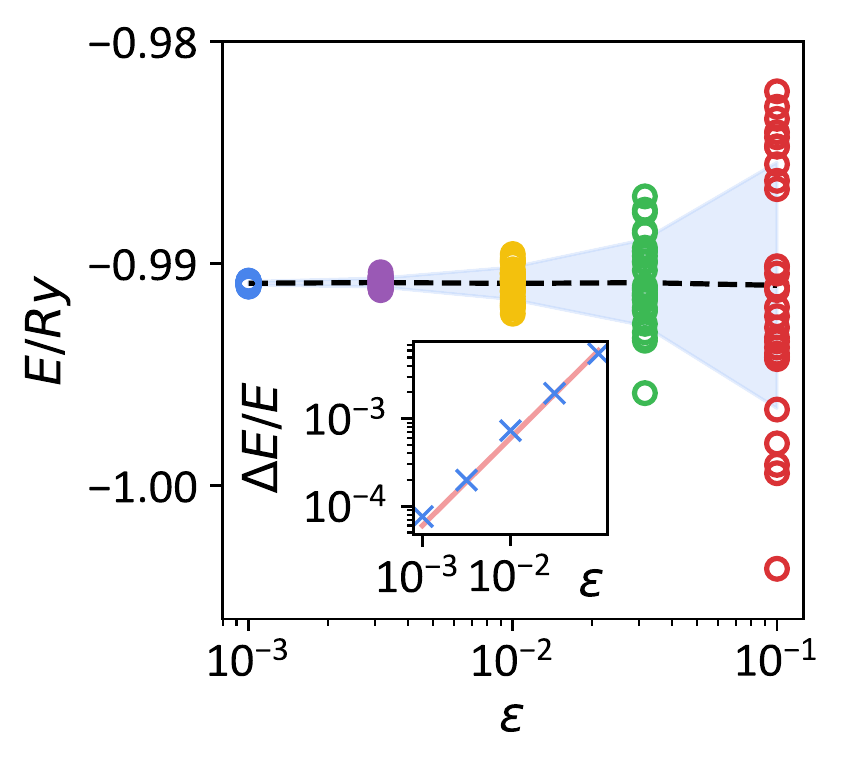}
  \caption{Ground-state energy of the 3D Hydrogen-like Hamiltonian for increasing values of the average normalized error of the nuclear potential, $\varepsilon$. 30 random iterations are considered for each choice of $\varepsilon$, showing the average value (dashed line) and standard deviation (coloured region). Inset: the relative error in the resulting energy follows a linear scaling with the normalized nuclear potential error, $\Delta E/E\approx 0.06 \varepsilon$ (red line). \emph{Parameters:} $N=60$, $t_f/V_0=1$.}
 \label{fig:70HydrogenGroundError}
\end{figure}

Beyond this error caused by discretization and finite size effects, in Fig.~\ref{fig:70HydrogenGroundError} we further benchmark energy deviations in the Hydrogen ground-sate caused by a relative normal error $\varepsilon$ in the induced nuclear potential (see Sec.~\ref{subsec:nuclear}). For lattice size $N=60$, we observe that $\varepsilon\sim 0.1$ (compatible with $R_f=2$ in Fig.~\ref{fig:holo_bigger}), already provides an accuracy in the retrieved energy of order $1\%$.

\section{Simulating electron repulsion in optical lattices}
\label{sec:qcelectron}

In this section, we explain how to obtain the interacting part of the quantum chemistry Hamiltonian as given by Eq.~\ref{subeq:rep}. This is the most complicated part of the simulation since it requires to describe long-range density-density interactions between ultra-cold fermionic atoms, whose interactions are typically local. That is, they only interact when their wavefunctions overlap significantly (i.e., same site). As proposed in Ref.~\cite{arguello2019analogue}, the key idea consists in using an auxiliary atomic species trapped together with the fermions such that the long-range interactions are effectively mediated by it.  For concreteness, we assume this auxiliary atom to be a boson, although this will not play a big role for the physics that will be discussed along this manuscript. These auxiliary atoms need to be trapped in an optical lattice of similar wavelength~\footnote{It would be enough that the period of the auxiliary atom lattice is commensurate} than the one of the fermions, and it should be able to interact locally with the fermions through the following Hamiltonian:
\begin{align}
\hat{H}_{\mathrm{f}-\mathrm{aux}}=U \sum_{\jj,\sigma} \hat{f}^\dagger_{\jj,\sigma} \hat{f}_{\jj,\sigma} \hat{b}^\dagger_\jj \hat{b}_\jj\,,\label{eq:Hfaux}
\end{align}
where $\hat{b}^\dagger_\jj (\hat{b}_\jj)$ are the creation/destruction operators associated to the bosonic atoms at the $\jj$ site. We consider that the bosonic optical lattice can have a different size than the fermionic one, i.e., having $N_M$ lattice sites. These atoms will have an internal dynamics described by a Hamiltonian $\hat{H}_\mathrm{aux}$ that will depend on the particular optical lattice configuration chosen, and that will ultimately determine the effective interactions induced in the fermionic atoms. Thus, the idea consists in properly engineering $\hat{H}_\mathrm{aux}$ such that the effective fermionic interactions give rise to the desired pair-wise, Coulomb potential.

In what follows, in Section~\ref{subsec:formalism} we first explain the general formalism that we will use to analyze this problem. Unlike the proposal of Ref.~\cite{arguello2019analogue}, here we introduce two simplified setups that we analyze in Sections~\ref{subsec:atoms}-\ref{subsec:spin}, and that will result in slightly different repulsive potentials from the targetted one. The complete proposal of Ref.~\cite{arguello2019analogue} is discussed in Section~\ref{subsec:spincavity}, where we also numerically benchmark that the perturbative working conditions that were derived in Ref.~\cite{arguello2019analogue} are correct. The motivation for this incremental discussion is two-fold: first, it allows one to understand the role of all the ingredients required in the final proposal; second, even though the models discussed in Sections~\ref{subsec:atoms} and~\ref{subsec:spin} do not provide a fully scalable Coulomb-like interaction, they can be used as simpler, but still meaningful, experiments that can simulate chemistry-like behaviour and guide the way to the full proposal.

\begin{figure*}[tbp]
	\centering
	\includegraphics[width=0.8\linewidth]{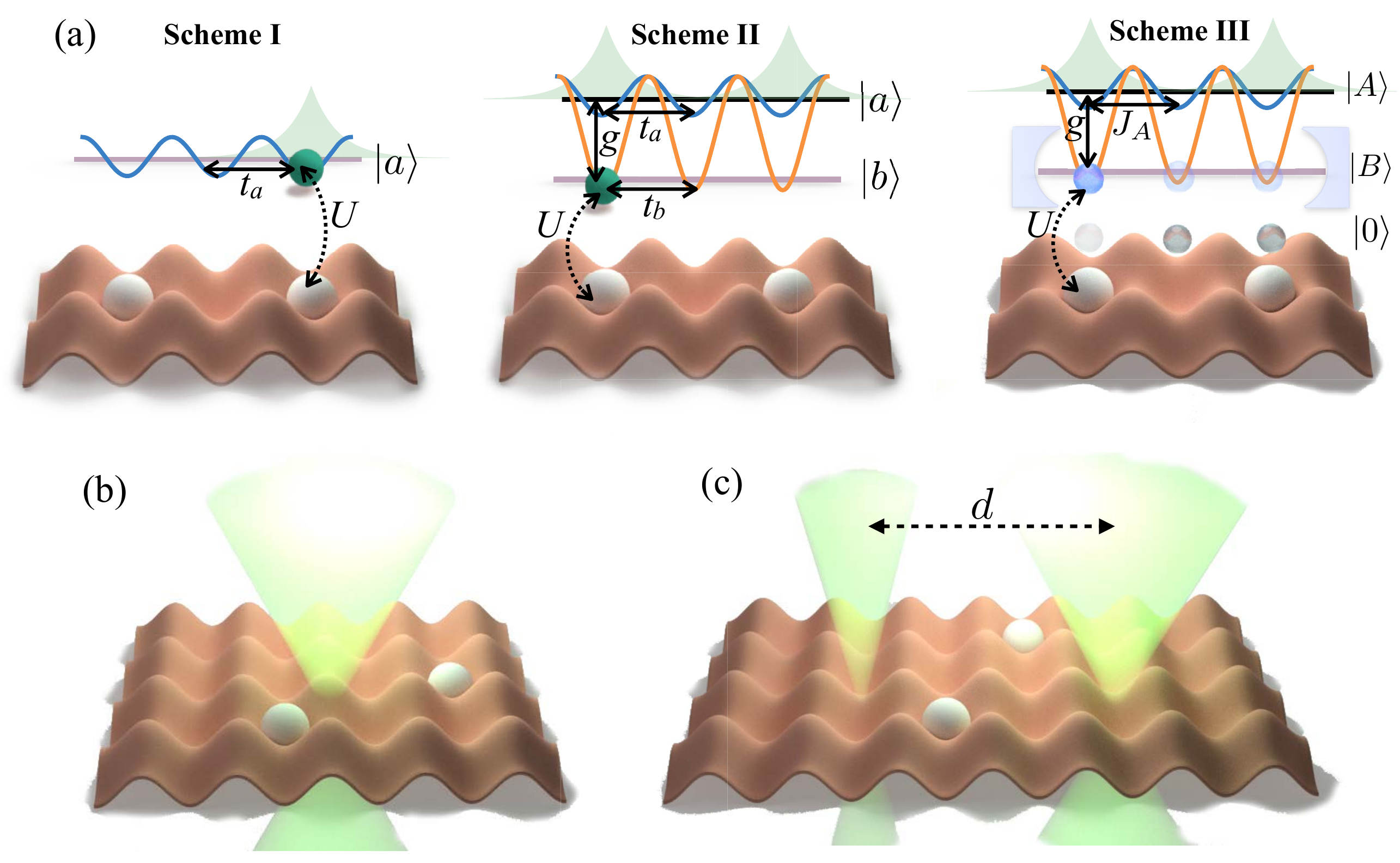}
	\caption{(a) Schemes presented in this work, to induce repulsion among fermionc cold atoms (white) trapped in a three-dimensional optical lattice (red, pictorically represented in a lower dimension for simplicity). In \textit{Scheme I} (Sec. \ref{subsec:atoms}), repulsion is induced by a single mediating atom (green) in a given internal state $\ket{a}$. This mediating atom tunnels with rate $t_a$, and experiences an on-site interaction with occupying the same lattice site as the fermionic atom. This induces a non-Coulomb and non-scalable repulsion for more than two fermionic atoms. In \textit{Scheme II} (Sec. \ref{subsec:spin}), one relies on a second internal level of the mediating atom to reach the Coulomb-like potential. Both levels $\ket{a}$ and $\ket{b}$ are coupled with rate $g$, and tunneling rate depends on their state.  In \textit{Scheme III} (Sec. \ref{subsec:spincavity}) we use a different approach to achieve the Coulomb repulsion among an arbitrary number of atoms. Here, rather than a single atom, the mediating species is in a Mott phase, with exactly one atom occupation per site (translucent gray). While these atoms are fixed, an atomic excitation (coloured blue) $\ket{A}$ can propagate to a neighboring site through magnetic exchange. They are also coupled with rate $g$ to another internal level $\ket{B}$, that is subject to a Raman-assisted cavity-mediated collective interaction, $J_c$, and a on-site repulsion $U$ with fermionic atoms. (b) Configurations corresponding to the simulation of atomic He, where a nuclear potential centered in the lattice is induced using holographic techniques~\ref{subsec:nuclear}, and two fermionic atoms trapped in the lattice play the role of two electrons. (c) The configuration is modified to simulate HeH$^+$, where one tailors the attraction due to to the distinct Hydrogen (left) and Helium (right) nuclear charges, separated, in this scheme, by $d/\text{a}=3$ sites. }
	\label{fig:scheme}
\end{figure*}

\subsection{General formalism}
\label{subsec:formalism}

Our approach to calculate the effective fermionic interactions will be based on the separation of energy scales between the fermionic dynamics ($\hat{H}_f$) and the rest ($\hat{H}_{f-\mathrm{aux}}+\hat{H}_\mathrm{aux}$). In particular, we will assume that $||\hat{H}_f||\ll ||\hat{H}_{f-\mathrm{aux}}+\hat{H}_\mathrm{aux}||$, such that we can consider the fermions fixed in the auxiliary atomic timescales. Thus, if we have $N_e$ electrons placed in positions $\lla{\jj}\equiv \jj_1,\dots,\jj_{N_e}$ we can make the following ansatz for the full atomic mixture wavefunction:
\begin{equation}
\label{eq:totBase}
\ket{\Psi_{f-\mathrm{aux}}\pa{\lla{\jj}}} = \ket{\jj_1,\dots,\jj_{N_e}}_f \otimes \ket{\varphi(\jj_1,\dots,\jj_{N_e})}_\mathrm{aux}\,.
\end{equation}
In this way, one can first solve the problem for the auxiliary atoms degrees of freedom within a fixed fermionic configuration $\{\jj_i\}_{i=1}^{N_e}$: 
\begin{align}
&\left[\hat{H}_{f-\mathrm{aux}}(\{\jj\})+\hat{H}_\mathrm{aux}\right]\ket{\varphi_m(\{\jj\}}_\mathrm{aux}=\nonumber\\
&=E_{m,\mathrm{aux}}(\{\jj\})\ket{\varphi_m(\{\jj\}}_\mathrm{aux}\,,\label{eq:aux}
\end{align}
where the index $m$ denotes the possible eigenstates within the same fermionic configuration, and where $\hat{H}_{f-\mathrm{aux}}(\{\jj\}$ reads:
\begin{align}
\hat{H}_{f-\mathrm{aux}}(\{\jj\}=U\sum_{\{\jj\}} \hat{b}^\dagger_{\jj_i} \hat{b}_{\jj_i}\,.
\end{align}
Note that $\sum_{\{\jj\}}$ indicates a sum now only over the fermionic positions.

Once the auxiliary atom problem of Eq.~\eqref{eq:aux} is solved, we divide the bosonic Hilbert space for each fermionic configuration distinguishing between the contribution of one of the eigenstates, $\ket{\varphi_s(\{\jj\}}_\mathrm{aux}$ with eigenenergy $E_{s,\mathrm{aux}}(\{\jj\})$, and the rest of states, that we label as $\ket{\varphi^\perp_{m}(\{\jj\}}_\mathrm{aux}$, with $m=1,\dots, N_M-1$. Then, one can calculate what is the effective fermionic Hamiltonian resulting from the dressing of such particular eigenstate $\ket{\varphi_s(\{\jj\}}_\mathrm{aux}$ by projecting in this space all possible fermionic configurations. The resulting effective Hamiltonian for the fermions reads
\begin{subequations}
\begin{align}
\hat{H}_\mathrm{eff}&=\sum_{\mathrm{all}\{\jj\}}\bra{\varphi_s(\{\jj\}}\left[\hat{H}_f+\hat{H}_{f-\mathrm{aux}}+\hat{H}_\mathrm{aux}\right]\ket{\varphi_s(\{\jj\}}_\mathrm{aux}\nonumber\\
&\approx -t_f \mathcal{F} \sum_{\mean{\ii,\jj},\sigma} \hat{f}^\dagger_{\ii,\sigma}\hat{f}_{\jj,\sigma}-\sum_{\alpha=1}^{N_n}\sum_{\jj,\sigma}\frac{Z_\alpha V_0}{|\jj-\RR_\alpha|}\hat{f}^\dagger_{\jj,\sigma}\hat{f}_{\jj,\sigma}+ \\
& +\sum_{\mathrm{all}\{\jj\}} E_{s,\mathrm{aux}}(\{\jj\}) \ket{\{\jj\}}\bra{\{\jj\}}\,,\label{eq:manyrep}
\end{align}
\end{subequations}
where we see how the auxiliary atomic state has two effects over the fermionic Hamiltonian.  First, it renormalizes the fermionic kinetic energy through the Franck-Condon coefficient:
	\begin{align}
	\label{eq:franckCondon}
	\mathcal{F}=\langle \varphi_s(\{\jj\}\ket{\varphi_s(\{\ii\}}_\mathrm{aux} \,,
	\end{align}
	that is the overlap between the bosonic states for two fermionic configurations $\{\jj\},\{\ii\}$ in which all the fermions have the same position, except one that is displaced to a nearest neighbour position. As we will see, $\mathcal{F}$ can be considered independent of the particular position occupied by the fermions. Since the only effect of this term is to renormalize the kinetic energy, in what follows we can assume $t_f \mathcal{F}\rightarrow t_f$,  and still write the single particle part of $\hat{H}_\mathrm{eff}$ like the $\hat{H}_f$ of Eq.~\eqref{eq:Hf3}.
	
 Second, and more importantly, a position-dependent energy-term which, in principle, depends on all the fermion positions, being therefore $2N_e$-body operator. However, when $E_{s,\mathrm{aux}}(\{\jj\})$ can be written as a sum of pairwise contributions, i.e., 
	\begin{align}
	 E_{s,\mathrm{aux}}(\{\jj\})=\sum_{\substack{m,n=1,\\ m\neq n}}^{N_e} V(\jj_m-\jj_n)\,,
	\end{align}
	the term of Eq.~\eqref{eq:manyrep} reduces to a density-density operator
	\begin{align}
	\hat{H}_\mathrm{eff}\approx \hat{H}_f+ \sum_{\substack{\sigma | m,n=1,\\ m\neq n}}^{N_e}  V(\jj_m-\jj_n) \hat{f}^\dagger_{\jj_m,\sigma}\hat{f}^\dagger_{\jj_n,\sigma}\hat{f}_{\jj_m,\sigma}\hat{f}_{\jj_n,\sigma}\,,\label{eq:Heff}
	\end{align}
	that will mimic that of the quantum chemistry Hamiltonian of Eq.~\eqref{subeq:rep} if $V(\jj_m-\jj_n)= V_0\text{a}/|\jj_m-\jj_n|$, with $V_0>0$ in order to be repulsive.

To conclude, let us summarize then the conditions to achieve the fermionic repulsion:
\begin{itemize}
	\item There should be one eigenstate $\ket{\varphi_s(\{\jj\}}_\mathrm{aux}$ from the Hamiltonian of Eq.~\eqref{eq:aux}, whose energy can be written as:
	\begin{align}
	\label{eq:cond1}
     E_{s,\mathrm{aux}}(\{\jj\})=\sum_{\substack{m,n=1,\\ m\neq n}}^{N_e} \frac{V_0 \text{a}}{|\jj_m-\jj_n|}\,,
	\end{align}
	where $V_0>0$ determines the strength of the effective repulsion between the electrons.
	
	\item For self-consistency, $\ket{\varphi_s(\{\jj\}}_\mathrm{aux}$ needs to be the dominant state of the Hilbert space of the auxiliary atoms dressing the fermionic configuration $\{\jj\}$, so that the total state writes as $\ket{\Psi} = \sum_{\lla{\jj}} \psi(\lla{\jj}) \ket{\{\jj\}}_f \ket{\varphi_s(\{\jj\}}_\mathrm{aux}$. For consistency, the different parts of the Hamiltonian ($\hat{H}_f$, $\hat{H}\st{f,aux}$,$\hat{H}\st{aux}$) should not couple significantly these state to the orthogonal ones $\ket{\varphi^\perp_m(\{\rr\}}_\mathrm{aux}$ of any given fermionic configuration. This means that, if any of the Hamiltonian parts $\hat{H}_\alpha$ connect $\ket{\varphi_s(\{\jj\}}_\mathrm{aux}$ to an state $\ket{\varphi^\perp_m(\{\rr\}}_\mathrm{aux}$, the transition should be prevented by a large enough energy gap between them, denoted by $\Delta_{m,\rr}$. Like this, we can upper-bound the error introduced by such couplings using perturbation theory:
	\begin{align}
	\label{eq:cond2a}
	 \varepsilon_\alpha=\sum_{\mathrm{all}\{\rr\}}\sum_{m}\left|\frac{_f\bra{\{\rr\}}_\mathrm{aux}\bra{\varphi^\perp_{m,\{\rr\}}}\hat{H}_\alpha\ket{\Psi}}{\Delta_{m,\rr}}\right|^2\,,
	\end{align}
    which should of course satisfy:
    \begin{align}
    \label{eq:cond2}
    \varepsilon_\alpha \ll 1\,,
    \end{align}
    for all $\alpha=\mathrm{f,f-aux,aux}$. This provides a second working condition for the dynamics to be governed by the effective fermionic repulsion Hamiltonian of Eq.~\eqref{eq:Heff}. 
\end{itemize}

In what follows, we will introduce sequentially the different schemes for the interaction of the mediating atoms in 3D, $\hat{H}_\mathrm{aux}$, until we obtain the desired repulsive, pair-wise, Coulomb potential between the fermionic atoms. For notation simplicity from now on, we will omit the fermionic spin degree of freedom in $\hat{f}_\jj$, but since the fermion-auxiliary atom interactions in Eq.~\eqref{eq:Hfaux} are assumed to be equal for both spin states, so will be the effective fermionic repulsion.

\subsection{Scheme I: Repulsion mediated by single atoms: non-Coulomb \& non-scalable}
\label{subsec:atoms}

Let us assume initially the simplest level configuration for the auxiliary atomic state, that is, it has only a single ground state level subject to an optical potential with the same geometry as the fermionic one, but with different amplitude [see Fig.~\ref{fig:scheme}(a)]. The resulting auxiliary Hamiltonian in this case reads as:
\begin{align}
\hat{H}\st{I,aux}=-t_b \sum_{\mean{\ii,\jj}} \hat b^\dagger_\ii \hat b_\jj\,,
\end{align}
where  $\hat b_\jj^{(\dagger)}$ represents the annihilation (creation) of auxiliary atoms at positions $\jj$, and $t_b$ their effective tunneling amplitude to the nearest neighbouring site. 

 Note that this Hamiltonian can be easily diagonalized in momentum space by introducing periodic boundary conditions, where $\hat{H}\st{I,aux}$ reads:
\begin{align}
\hat{H}\st{I,aux}=\sum_\kk \omega_\kk \hat{b}_\kk^\dagger \hat{b}_\kk\,
\end{align}
being $\hat{b}^{\dagger}_\kk=\frac{1}{\sqrt{N_M}}\sum_\jj \hat{b}_\jj\dg e^{i \kk\cdot\jj}$, and $\hat{b}_\kk=\frac{1}{\sqrt{N_M}}\sum_\jj \hat{b}_\jj e^{-i \kk\cdot\jj}$, the atomic creation and annihilation operators in momentum space, and
  $\omega_\kk=-2 t_b\sum_{\alpha=x,y,z}\cos(k_\alpha)$ their corresponding eigenenergies for a given momentum vector $\kk=(k_x,k_y,k_z)$, with $k_\alpha \text{a}\in \lla{ 2j\pi/N_M \text{ for }j=1\ldots N_M}$, $\alpha\in\lla{x,y,z}$, and $N_M$ the total number of sites of the auxiliary atom potential along one direction. 
  
  For the purpose of this section, we will focus on a single auxiliary atom living in the lattice. In this case, one can write an ansatz for the wavefunction of the auxiliary atom $\ket{\phi_m(\{\jj\})}_\mathrm{aux}=\sum_\kk \phi_{m,\kk}(\{\jj\}) \hat{b}_\kk^\dagger \ket{\mathrm{vac}}$ that can be used to find their corresponding eigenenergies:
  \begin{align}
  \label{eq:bosonModel1a}
  \left[U\sum_{\{\jj\}}\hat{b}^\dagger_\jj\hat{b}_\jj +\hat{H}\st{I,aux}\right]\ket{\phi_m(\{\jj\})}_\mathrm{aux}=E_m(\{\jj\}) \ket{\phi_m(\{\jj\})}_\mathrm{aux}\,.
  \end{align}
  
  In what follows, we analyze first the case of having a single fermion in the system where we will see the emergence of a bound state of the auxiliary atom around the fermionic position~\cite{Calajo2016,Shi2016,Shi2018}. Then, we will see how this bound state can mediate a repulsive interaction when two or more fermions are hopping in the lattice that, unfortunately, does not have the correct spatial dependence presented in Eq.~\eqref{eq:cond1}.

  \subsubsection*{Single fermion}
  
  If only a single fermion is present at the system at position $\jj_0$, then Eq.~\eqref{eq:bosonModel1a} leads to the following equation:
    \begin{equation}
  \label{eq:1boundstate}
      U^{-1}=\frac{1}{N_M} \sum_\kk \frac{1}{E\st{I,B}-\omega_\kk}\,,
  \end{equation}
  which has a bound-state solution for the auxiliary atom whose energy $E\st{I,B}$ lies above the scattering spectrum, i.e., $E\st{I,B}>6t_b$. Its associated wavefunction in the position representation, $\ket{\varphi_{B,\jj_0)}}=\sum_\rr \varphi_{B,\jj_0}(\rr)\, \hat b_\rr^\dagger$ reads as:
    \begin{equation}
  \label{eq:1boundwave}
     \varphi_{B,\jj_0}(\rr)=\frac{1}{\sqrt{\mathcal N_B}\,  N_M} \sum_\kk \frac{e^{i\kk\cdot(\rr-\jj_0)}}{E\st{I,B}-\omega_\kk}\,,
  \end{equation}
  with $\mathcal N_B=\frac{1}{N_M} \sum_\kk \frac{1}{\pa{E_B-\omega_\kk}^2}$. Taking the continuum limit, $N_M\rightarrow \infty$, to replace the summation by an integral, and making a quadratic expansion of the energy dispersion around the band-edges (see Appendix \ref{ap:perturbation}), one can obtain an analytical expression for the wavefunction that reads as:
  \begin{equation}
  \label{eq:mod1wave}
       \varphi_{B,\jj_0}(\rr)\propto\frac{e^{-\abs{\rr-\jj_0}/L\st{I}}}{r}\,.
  \end{equation}
  That is, a Yukawa-type localization around the fermionic position $\jj_0$ with a localization length given by $L\st{I}=\text{a}\sqrt{E\st{I,B}/t_b-6}$ which, to leading order in $t_b/U$, reads as (see Appendix \ref{ap:perturbation}):
  \begin{equation}
  \label{eq:lexp}
        L\st{I}=\text{a}\co{3.176-4\pi t_b/U}^{-1}\,.
  \end{equation}
\begin{figure}[tbp]
	\centering
	\includegraphics[width=0.7\linewidth]{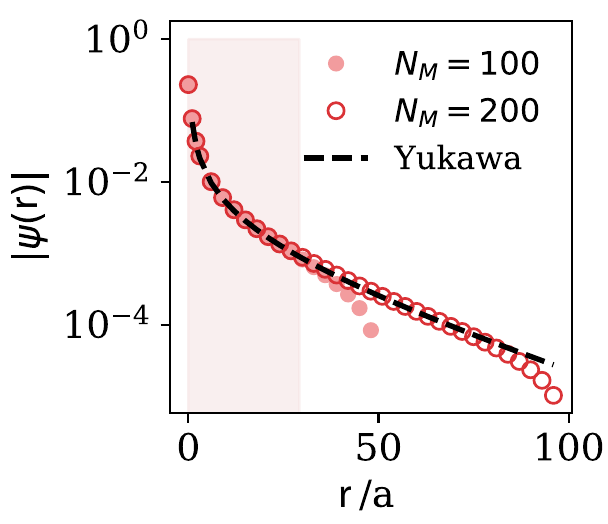}
 	\caption{Radial wavefunction of the mediating auxiliary atom placed at $\jj_0/\text{a}=\fl{N_M^{1/3}/2}[1,1,1]$ for lattice size $N_M=100^3,200^3$ and $U=4t_b$. Displacements are studied in the $x$-axis, $\jj=\jj_0+[r,0,0]$. Dashed line follows the Yukawa potential in Eq.~\eqref{eq:mod1wave}. Shaded region indicates the region $r/\text{a}\leq L\st{I}$ where Coulomb potential applies. Here, $L_I$ is calculated using Eq.~\eqref{eq:lexp}.}
	\label{fig:checkspatialScalings}
\end{figure}

Interestingly, this localization length can be tuned with the experimental parameters, i.e., changing $U/t_b$, and be made very large. In particular, the bound state can display a $1/r$ shape over the whole fermionic lattice as long as $L\st{I}\gg N^{1/3}$.

These analytical formulae can be numerically benchmarked by solving Eq.~\eqref{eq:1boundwave} for the case of a single fermion and a finite lattice size. This is done in Fig.~\ref{fig:checkspatialScalings} where we plot the spatial dependence of the numerically obtained wavefunction for two different system sizes: $N_M=100$ and $N_M=200$, represented with filled and empty circles, respectively, together with the Yukawa shape (in dashed black) predicted from Eq.~\eqref{eq:mod1wave}. We have chosen $U/t_b=4$ such that the expected length is $L\st{I}/\text{a}\approx29$, indicated by the shaded red region of the figure. From this figure we can extract two conclusions: first, the spatial wavefunction displays, as expected from Eq.~\eqref{eq:mod1wave}, an approximate $1/r$ decay for short distances, i.e., $r<L\st{I}$ (shaded red area). Second, for larger distances the spatial wavefunction follows the Yukawa shape of Eq.~\eqref{eq:mod1wave} until it becomes closer to the border. Thus, to observe the $1/r$ decay for the whole fermionic space we require that $N^{1/3}\ll L\st{I}\ll N_M^{1/3}$.

An additional condition comes from reducing the coupling to non-orthogonal states, i.e., the condition of Eq.~\eqref{eq:cond2}. In this case, only $\hat{H}_f$ contributes as follows:
\begin{align}
\label{eq:errorM1}
\varepsilon_f&=\sum_m \abs{\frac{t_f \braket{\varphi^\perp_{m,\jj_0+1}}{\varphi_{B,\jj_0}}}{\Delta_{m,\jj_0+1,\jj_0}}}^2\leq \frac{t_f^2}{\mathcal N_B N_M} \sum_{\kk} \frac{1}{\pa{E\st{I,B}-\omega_\kk}^4}
\end{align}
such that the condition $\varepsilon_f\ll1$ translates into:
\begin{align}
\label{eq:model1tftb}
t_f/t_b\ll \pa{a/L\st{I}}^2\,,
\end{align}
providing the working condition that guarantees the separation of energy scales between the fermionic and auxiliary atom dynamics. Here, $\jj_0+1$ denotes a nearest-neighbor of $\jj_0$, and we have made use of the calculations in Appendix  \ref{ap:perturbation}. This energy separation guarantees that the auxiliary atom will immediately follow the fermion as it hops through the lattice. As we already explained in the previous section, this auxiliary atom dressing renormalizes as well the fermion hopping by the Franck-Condon coefficient (see Eq.~\eqref{eq:franckCondon}). For the nearest-neighbour hoppings, that are the only non-negligible ones in this case, this coefficient reads
\begin{align}
\label{eq:f1}
 \mathcal{F}\st{I}= \braket{\varphi_{B,\jj_0+1}}{\varphi_{B,\jj_0}}\approx  e^{-\text{a}/L\st{I}}
\end{align}
so that the fermionic hopping is less affected the more delocalized the auxiliary atom wavefunction is.
 
    \subsubsection*{Two fermions}
    Let us now explain what occurs in the case where two fermions are placed at positions $\{\jj_1,\jj_2\}$. Solving the time-independent Schr\"odinger equation Eq.~\eqref{eq:bosonModel1a},
    one can find that now it has two, not one, bound-state solutions, i.e., with energies $E_{\pm
    } (\{\jj\})>6t_b$, and whose wavefunction in momentum space reads:
    \begin{equation}
    \label{eq:wavefunction2}
    \phi_{\pm,\jj}(\kk) \propto \frac{e^{-i\kk \jj_{1}}\pm e^{-i\kk \jj_{2}}}{E_{\pm
    	} (\{\jj\})-\omega_{\kk}}\,.
    \end{equation}
    
    When transforming these expressions into real space, one can see that they correspond to the combination of excitations states localized around the fermionic positions $\{\jj\}$. However, as explained in the previous section (see Eq.~\eqref{eq:cond1}), what governs the effective induced interaction between fermions is the spatial dependence of the eigenenergies $E_{\pm
    } (\{\jj\})$, which is given by,
    \begin{equation}
    \label{eq:2boundState}
    U^{-1}=\frac{1}{N_M}\sum_{\kk}\frac{1 \pm e^{i\kk\cdot\jj_{12}}}{E_{\pm}(\{\jj\})-\omega_{\kk}}\,,
    \end{equation}
     where $\jj_{12}=\jj_1-\jj_2$. The shape of this energy dependence depends on both the symmetric/antisymmetric character of the wavefunction, and whether the solution is found above/below the scattering spectrum ($\omega_\kk$), that can be tuned by modifying $U/t_b$. By numerical inspection, we observe that to obtain a repulsive interaction we must use the symmetric state and tune the parameters such that $ E_{+}(\{\jj\})>\omega_\kk$. In that case, it can be shown how the energy of the symmetric state can be written as
\begin{equation}
    E_{+}(\{\jj\})\approx E\st{I,B} + V\st{I}(\jj_{12})
\end{equation}
where $E\st{I,B}$ is the bound-state energy of a single fermion, and where the spatial dependence $V\st{I}(\jj_{12})$ is given by:
\begin{equation}
    V\st{I}(\jj_{12})\approx \frac{1}{\mathcal{N}_B N_M} \sum_\kk \frac{e^{i\kk \jj_{12}}}{E\st{I,B}-\omega_\kk}\,,
\end{equation}
that is the term that induces a position-dependent interaction between the fermions (see Eq.~\eqref{eq:cond1}). Note as well the similarity between $V\st{I}(\jj_{12})$ and the bound-state wavefunction of the single-fermion case (Eq.~\eqref{eq:1boundwave}). Thus, we can also take the continuum limit of this expression to transform the sums into integrals and make a parabolic expansion of $\omega_\kk$ to obtain an analytical formula of $V\st{I}(\jj_{12})$. In the long-distance limit, that is, when $|\jj_{12}|\gg L\st{I}$, the potential shows the same Yukawa shape:
\begin{align}
\label{eq:short1}
V_{\text{I},>}(\jj_{12})=V_{\text{I},>}(|\jj_{12}|)&\approx\frac{2\text{a}^2 t_b}{|\jj_{12}|\cdot L\st{I}} e^{-|\jj_{12}|/L\st{I}}\,.
\end{align}

Unfortunately, in the opposite limit, i.e., $|\jj_{12}|\ll L\st{I}$, where the shape should display the desired $1/|\jj_{12}|$ shape, Eq.~\eqref{eq:2boundState} induces an additional correction which yields (see Appendix \ref{ap:perturbation})
\begin{align}
\label{eq:long1}
V_{\text{I},<}(\jj_{12})/t_b&\approx \frac{0.322 \text{a}^2}{|\jj_{12}|^2}+\frac{0.724 \text{a}^2}{|\jj_{12}|\cdot L\st{I}}\,.
\end{align}

\begin{figure}[tbp]
	\centering
	\includegraphics[width=1\linewidth]{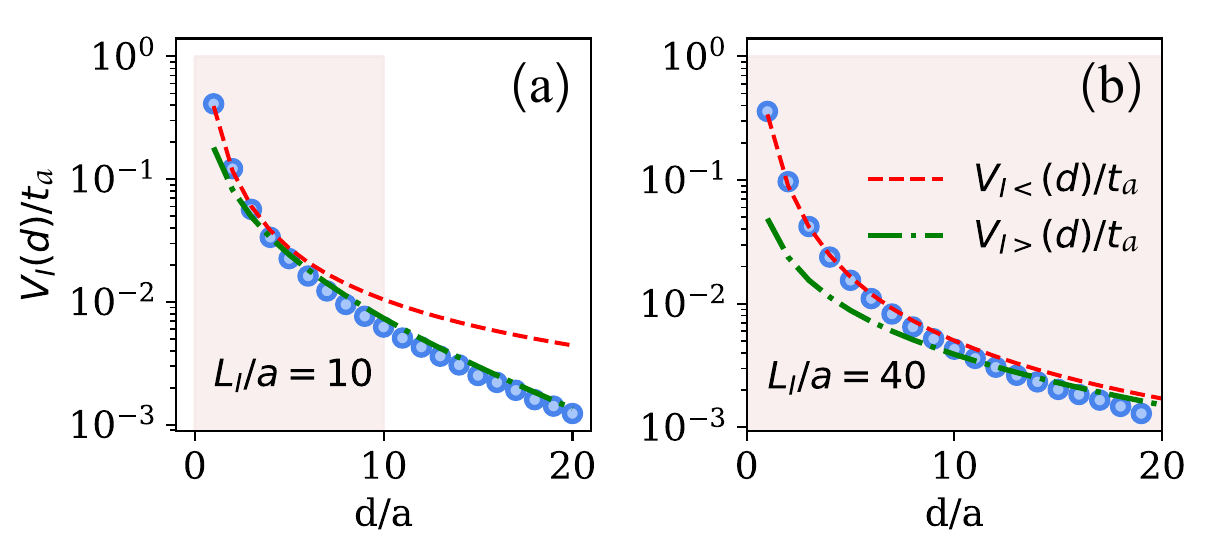}
	\caption{Comparison between the numerical calculation of $V\st{I}(d)$ obtained from equating Eqs.~\eqref{eq:1boundstate} and \eqref{eq:2boundState} (round markers), as compared to the analytical expansions \eqref{eq:short1} (dashed lined) and \eqref{eq:long1} (dotted-dashed line), valid in the regimes $d/L\st{I}\ll 1$ (coloured region) and  $d/L\st{I}\gg1$, respectively. \emph{Parameters:} $N_M=120^3$. }
	\label{fig:checkScalings}
\end{figure}

These analytical expressions are numerically benchmarked by solving the bosonic Hamiltonian \eqref{eq:bosonModel1a} in a finite system for two fermions separated by an increasing number of sites, and two different values $L\st{I}$, as shown in Fig.~\ref{fig:checkScalings}. There, we observe how the energy spatial decay never displays the desired $1/|\jj_{12}|$ scaling but rather the $1/|\jj_{12}|^2$ predicted by Eq.~\eqref{eq:long1}. The intuition behind this limitation is that we do not have enough tunable parameters since $U/t_b$ controls both the strength and the range of the interaction ($L\st{I}$). Thus, when $L\st{I}$ is tuned to be large enough, the correction to the energy $E_{\pm}(\{\jj\})$ is so strong that it induces a different spatial dependence from the $1/r$ shape. Let us also mention here that when $t_b>0$, there is an additional checkerboard phase pattern in the spatial dependence $V\st{I}(\jj_{12})$ that appears because the closer energy modes of the upper band-edge of $\omega_\kk$ have $\pm \pi$-momenta. Therefore, if one wants the fermion not to be sensitive to it, it is needed that the periodicity of the auxiliary atom lattice is half the one of the fermions. Another option consists on working with an excited energy band that shows $t_b<0$~\cite{muller07b}, so that this checkerboard phase pattern does not appear.

For completeness, let us also mention here that as the two fermions separate, the auxiliary atom wavefunction approximates a superposition of the single-boson density~ of Eq.~\eqref{eq:1boundwave} centered at each position, such that the Franck-Condon coefficient of Eq.~\eqref{eq:franckCondon} approximates as
$
\mathcal{F}\st{I}\approx 0.5\pa{1+e^{-\text{a}/L\st{I}}}\,.$ Additionally to the error in Eq.~\eqref{eq:errorM1} caused by the coupling to states in the band due to $\hat{H}_f$, the condition on $t_f/t_b$ derived for the single fermion case now includes an additional contribution given by Eq.~\eqref{eq:cond2} due to the coupling to the antisymmetric bound-state. This additional contribution reads as,
\begin{align}
\label{eq:errorM1b}
\abs{\frac{t_f \braket{\varphi_{-\pa{\jj_1+1,\jj_2}}}{\varphi_{+\pa{\jj_1,\jj_2}}}}{E_+\pa{\jj_1,\jj_2}-E_-\pa{\jj_1+1,\jj_2}}}^2\approx \abs{\frac{t_f\pa{1-\mathcal{F}\st{I}}}{4V\st{I}(d)}}^2\ll 1\,,
\end{align}
where, for $d/\text{a}\gg1$, we have approximated $\braket{\varphi_-\pa{\jj_1+1,\jj_2}}{\varphi_+\pa{\jj_1,\jj_2}}\approx 0.5\pa{1-\mathcal{F}\st{I}}$. From the definition of the Franck-Condon coefficient~\eqref{eq:f1}, in the limit $L\st{I}/\text{a}\gg1$, this can be approximated as $0.5 \text{a}/L\st{I}$. We observe that since the gap between the symmetric and antisymmetric state is given by $2V\st{I}(|\jj_{12}|)$, the condition becomes more demanding as the two fermions separate, since $V\st{I}(|\jj_{12}|\rightarrow \infty)\to 0$.

Ensuring that the symmetry of state is preserved irrespective of the fermionic positions will be one of the main motivations to introduce the cavity-assisted hoppings required for the model discussed in Section~\ref{subsec:spincavity}.

\subsubsection*{More than two fermions}

\begin{figure}[tbp]
	\centering
	\includegraphics[width=1\linewidth]{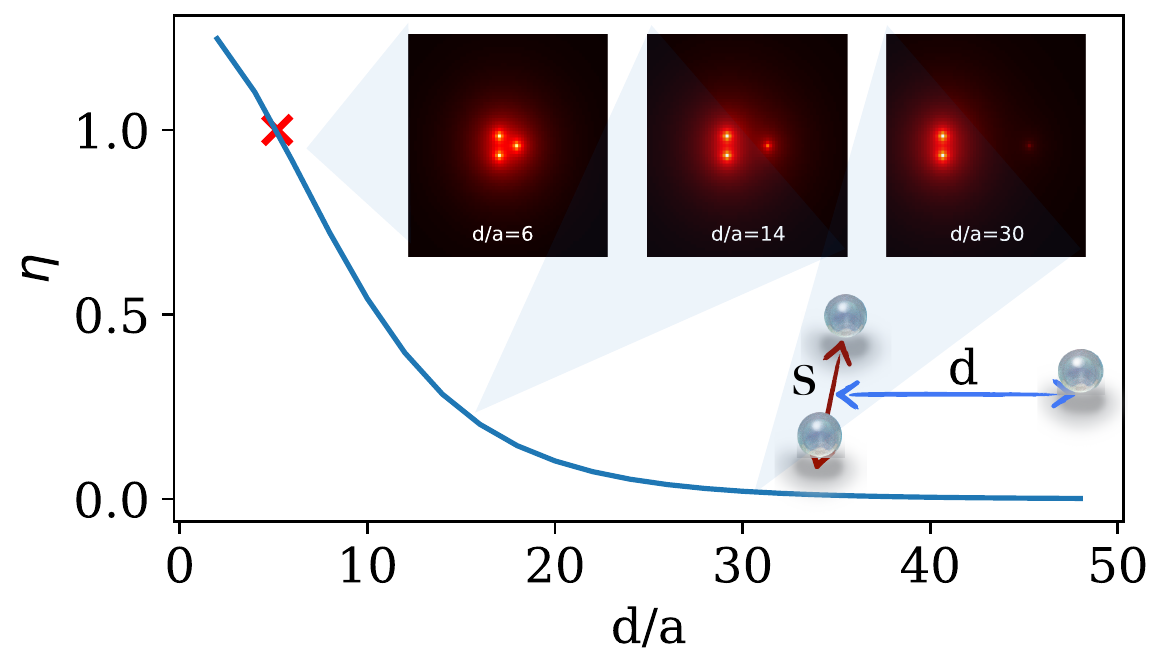}
	\caption{Asymmetry between the population of the bound state of Hamiltonian \label{eq:bosonModel1} for a configuration of three simulating atoms. Here we choose an isosceles distribution, of fixed base $s/\text{a}$, and variable height $d/\text{a}$ (see inserted scheme). Continuous line follow the ground state obtained with exact diagonalization for different values of $d/a$. $\eta$ denotes the ratio between the population of the bath in the apex site, and one of the vertices of the base. A value 1, indicating the desired symmetric superposition between the three atoms, is only achieved in the equilateral configuration (red cross). Panels show axial cuts of the population in bath $b$ for geometrical configurations $d/\text{a}=6,14,$ and $30$ (from left to right). Qualitatively, one observes that the symmetric superposition at the three vertices is less enforced as triangle sides become more unequal. \textit{(Parameters:} $L\st{I}/\text{a}=10$, $s/\text{a}=6$, $N_M=120^3$). }
	\label{fig:2cavity_no}
\end{figure}

Although we already showed in the previous section that this auxiliary atom configuration will not be able to deliver the desired Coulomb potential for two fermions, let us here consider the general situation with $N_f$ fermions to see that an additional complication arises, that is, that the eigenenergy $E_+(\{\jj\})$ does not correspond to a pair-wise sum. Instead, the auxiliary atomic excitation tends to localize more strongly around those fermions closer to each other, making the proposal non-scalable. To illustrate this effect, in Fig.~\ref{fig:2cavity_no} we plot an example of a numerically calculated energy $E_{+}(\{\jj\})$ when three fermions are placed in a triangular disposition and move the distance of one of them such that it goes from an equilateral configuration to an isosceles one. We plot the ratio between the population in the fermionic sites at the apex of the triangle, compared to one of the positions of the base ($\eta$ in the figure). There, we observe that the population only becomes equal in the equilateral superposition.

Having identified the problems with this simple auxiliary atom configuration, in the next subsections we will show how by adding complexity to the internal dynamics of the auxiliary atom, one can solve these problems.

\subsection{Scheme II: Repulsion mediated by atoms subject to state-dependent potentials: Coulomb but non-scalable}
\label{subsec:spin}
One of the problems of the previous proposal is the impossibility of independently tuning the strength and range of the interactions, since there is only a single tunable parameter ($U/t_b$). Here, we will show how to harness the latest advances in state-dependent optical lattices~\cite{hofrichter16a,snigirev17a,riegger18a,krinner18a,heinz20a} to gain that tunability.

\begin{figure}[tbp]
	\centering
	\includegraphics[width=0.8\linewidth]{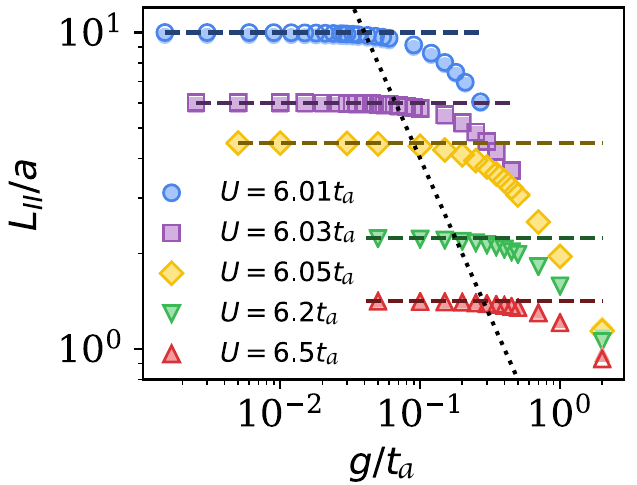}
	\caption{
	Effective Yukawa-type length obtained from fitting the exponentially decaying wavefunction in bath $a$ (shaded marker), and the state energy, $\pa{E\st{II,B}/t_a - 6}^{-1/2}$, (contoured marker) for different values of on-site interaction $U/t_a$. Dashed lines follow the leading-order approximation $L\st{II}^{(0)}/\text{a}= \pa{E\st{II,B}^{(0)}/t_a-6}^{-1/2}$, and the dotted line indicates the critical coupling strength $g/t_a$ when second-order corrections in the population $a$-modes are 1$\%$ larger than the leading order. Here, $N_M=100^3$.}
	\label{fig:32checkCondition}
\end{figure}

The idea consists of assuming that one can engineer two very different potentials for two long-lived states of the auxiliary atoms that we label as $a$ and $b$ (see Fig.~\ref{fig:scheme}(b) for a scheme), such that when the atoms are in $b$, they tunnel at a much slower rate, $t_b$, than when they are in $a$, i.e.,  $t_a\gg t_b$. These states can be either the hyperfine states of an Alkali specie, or the metastable excited states of Alkaline-Earth ones. What is important is that these states can be coherently coupled either  through a two-photon Raman transition or a direct one with effective coupling amplitude $g$ and detuning $\Delta$. Like this, the global internal dynamics for the auxiliary atom is described by the following Hamiltonian: 
\begin{align}
\label{eq:2levelHamiltonian}
\begin{split}
\hat H\st{II,aux} =&\Delta \sum_\jj  \hat{b}_\jj^\dagger \hat{b}_\jj -t_a \sum_{\langle \ii , \jj\rangle }\hat{a}_\ii^\dagger \hat{a}_\jj+ g\sum_\jj (\hat{b}_\jj^\dagger \hat{a}_\jj +\hc)\\
&-t_b \sum_{\langle \ii , \jj\rangle }\hat{b}_\ii^\dagger \hat{b}_\jj 
\,.
\end{split}
\end{align}
Using this Hamiltonian, one can solve again Eq.~\eqref{eq:bosonModel1a} for two fermions in a configuration $\{\jj\}$, but now replacing $H\st{I,aux}\rightarrow H\st{II,aux}$. One can write the following ansatz for the auxiliary atom wavefunction:
\begin{align}
\ket{\phi\st{II,m}(\{\jj\})}_\mathrm{aux}=\sum_\kk\left(\phi^a_{m,\lla\jj}(\kk) \hat{a}_\kk^\dagger+\phi^b_{m,\lla{\jj}}(\kk) \hat{b}_\kk^\dagger\right) \ket{\mathrm{vac}}\,.
\end{align}
Under these conditions, we find that there is again a symmetric bound state in bath $b$ localized around the fermions, whose associated eigenenergy $E\st{II,+}(\jj)$ leads to repulsive spatially-dependent interactions. Since $t_b\ll t_a$, the spatial dependence is dominated by the hopping in the $a$-bath. In order to obtain an analytical expression for $E\st{II,+}(\jj)$, we will further assume that $g\ll t_a$ and that $t_b=0$. Note that even if one takes originally $t_b= 0$, one does still obtain an effective tunnelling through the $a$ bath given by $t_b\approx g^2t_a/\Delta^2$, that we will neglect to get the analytical expression. These assumptions allow us to obtain $E\st{II,+}(\jj)$ using second-order perturbation theory, which yields: 
\begin{align}
 \label{eq:e2b}
E\st{II,+}(\{\jj\})&\approx E^{(2)}\st{II,B}+\frac{g^2}{N_M} \sum_{\kk} \frac{e^{i\kk\cdot\jj_{12}}} {E\st{II,B}^{(0)}-\omega_{\text{II},\kk}}\,,
\end{align}
where $\omega_{\text{II},\kk}=-2 t_a\sum_{\alpha=x,y,z}\cos(k_\alpha)$ is the energy-dispersion ruling the propagation of the $a$ modes, and $E\st{II,B}^{(0)}=U+\Delta$, $E\st{II,B}^{(2)}=E\st{II,B}^{(0)}+ \frac{g^2}{\mathcal{N}}\sum_{\kk} \frac{1} {E\st{II,B}^{(0)}-\omega_{\text{II},\kk}}$ are the bound-state energies for the single-fermion case in this atomic configuration calculated to $0$-th/$2$-nd order, respectively (see Appendix~\ref{ap:perturbation} for more details on the calculation). As we did for $E\st{I,+}(\{\jj\})$ one can obtain a formula for the spatial dependence taking the continuum limit to transform the sums into integrals, and expanding $\omega_\kk$ around its band-edges, yielding
\begin{align}
E\st{II,+}(\{\jj\})-E^{(2)}\st{II,B}&\approx V\st{II}\frac{\text{a} e^{-|\jj_{12}|/L^{(0)}\st{II}}}{|\jj_{12}|}\,,
\label{eq:e2b2}
\end{align}
with $V\st{II}=g^2/(4\pi t_a)$ being the strength of the repulsive interaction, and  $L\st{II}^{(0)}=\text{a}\cdot \pa{E\st{II,B}^{(0)}/t_a-6}^{-1/2}$ its range calculated using the $0$-th order energy. The latter can also be calculated exactly obtaining a value $L\st{II}$ that should ideally satisfy $L\st{II}\approx L\st{II}^{(0)}$ (see Fig.~\ref{fig:32checkCondition} and the discussion around it). From Eq.~\eqref{eq:e2b2} we can already see that this atomic configuration solves one of the problems of the previous proposal of section~\ref{subsec:atoms}, that is, that now one can tune independently the strength $V\st{II}$ and  its range $L\st{II}$. This enables going to a regime where $L\st{II}$ is bigger than the fermionic system size, i.e., $L\st{II}\gg N^{1/3}$, while still keeping the $1/r$-dependence such that the two-fermion repulsion has a truly Coulomb-like shape in all space. 

 Now, let us see the working conditions, based on the discussion around Eq.~\eqref{eq:cond2}, where this effective repulsion works.

\begin{itemize}
	\item Let us first bound the corrections introduced by the fermion hopping Hamiltonian $\hat{H}_f$. Focusing on the two-fermion case, these contributions are:
	\begin{align}
	\varepsilon_f&=\abs{\frac{t_f}{U}}^2+\abs{\frac{\text{a}}{L\st{II}}\frac{t_f }{4V\st{II}(d)}}^2\ll 1
	\end{align}
	where the first term corresponds to the coupling to states $\hat b_\jj\dg \ket{\text{vac}}$ in positions not occupied by a fermion, and the second term corresponds to the antysimmetric state whose population in level $b$ is approximately $\pa{\hat b_{\jj_1}\dg-\hat b_{\jj_2}\dg}/\sqrt{2} \ket{\text{vac}}$, analogously to Eq.~\eqref{eq:errorM1b}. As it occurred in the previous model, ensuring the right symmetry for the mediating state becomes more demanding as the two fermions separate. From the definition of the Bohr-radius~\eqref{eq:Bohr}, larger orbital sizes require to increase the effective length of the Yukawa potential so that, in the worst-case-scenario where the fermions are maximally separated, $\pa{\text{a}_0/\text{a}}\pa{N/L\st{II}}\ll 1$ is still satisfied. 
	
	\item The correction introduced by $H\st{II,aux}$ can be bounded by (see Appendix~\ref{ap:perturbation} for details):
	\begin{align}
	\label{eq:model2Aux}
	\varepsilon_\mathrm{aux}\leq  \frac{V\st{II}L\st{II}}{t_a\text{a}} \ll 1 \,,
	\end{align}
	that guarantees that the population in the $a$ modes remains small, such that the second-order expansion used in Eq.~\eqref{eq:e2b} holds.
	
	\item Besides, as aforementioned, it is desirable that the localization length $L\st{II}$ is independent on the particular fermionic configuration. However, by solving numerically Eq.~\eqref{eq:bosonModel1a} with $H\st{II,aux}$ for a single fermion, we find that the length of the bound state that will afterwards mediate the interaction can depend on the ratio $g/t_a$, and thus on $V\st{II}(d)$. This is shown explicitly in Fig.~\ref{fig:32checkCondition} where we plot the $L\st{II}$ obtained by a numerical fitting of the bound-state shape as a function of $g/t_a$ and for several $U/t_a$, and compare it with $L\st{II}^{(0)}$ (dashed black lines). There, we observe how indeed $L\st{II}^{(0)}$ matches well the numerically obtained value until a critical $g/t_a$ where it starts to deviate significantly. We numerically observe that $L\st{II}$ deviates significantly from $L\st{II}^{(0)}$, when the population in $a$-mode deviates from its first-order expansion terms (in dashed black). Using that intuition, we can then estimate the conditions for the $L\st{II}$-independence of parameters by imposing that the higher-order terms in the $a$-modes are smaller than the first order ones, which yields the following inequality (see Appendix \ref{ap:perturbation})
	\begin{align}
	\label{eq:lIIcondition}
	V\st{II}/t_a \ll \pa{\text{a}/L\st{II}}^2
	\end{align}
	 
	 From an energy perspective, we see that this bound obtained from the population of atoms in level $ a$, dictates that the mediated repulsion, $E\st{II,+}(\{\jj\})-E^{(2)}\st{II,B}$, needs to be smaller than the energy-gap, $E^{(2)}\st{II,B}/t_a-6$, defining $L\st{II}^{(0)}$. This condition also ensures that the higher-order corrections to the bound-state energy dependent on the fermionic configuration can be neglected.
\end{itemize}

Under these conditions, this experimental setup allows us to simulate faithfully a quantum chemistry interaction for two-electron (fermion) problems. Unfortunately, this proposal inherits the same problems of scalability than the previous one: when more than two fermions are present, the bound state tends to localize more strongly in the position of the closest ones (remember Fig.~\ref{fig:2cavity_no}), and $E\st{II,+}$ can not be written as a pairwise potential. 

\subsection{Scheme III: Repulsion mediated by atomic spin excitations and cavity assisted transitions}
\label{subsec:spincavity}

For completeness of this manuscript, we finally review here the proposal introduced in Ref.~\cite{arguello2019analogue} with all the ingredients required to obtain the repulsive, pair-wise, $1/r$, potential needed for quantum chemistry simulation. The goal is two-fold: on the one hand, the previous analysis of the simplified setups will allow us for a more intuitive understanding of the role of the different elements. On the other hand, we will numerically benchmark through exact calculations the working conditions of the simulator derived perturbatively in Ref.~\cite{arguello2019analogue}.

This proposal requires (see Fig.~\ref{fig:scheme}(a)):
\begin{itemize}
	\item Three long-lived states that we label as $a,b,c$, subject to different state-dependent potentials, such that they can only hop when they are in the $a$ state.
	
	\item The auxiliary atoms should be initialized in a Mott-insulating state $\ket{\mathrm{Mott}}=\prod_{\ii}\hat{c}^\dagger_\ii\ket{\mathrm{vac}}$ with unit filling.  Like this, instead of working with atomic excitations directly like we did in the previous two subsections, the second-quantized operators $\hat{A}_\jj,\hat{B}_\jj$ will denote single-spin excitations over the Mott-state, i.e., 
	\begin{align}
	\hat{A}_\jj^\dagger/\hat{B}_\jj^\dagger \ket{\mathrm{Mott}}=\left(\prod_{\ii\neq \jj}\hat{c}^\dagger_\ii\right) \hat{a}^\dagger_\jj/\hat{b}^\dagger_\jj\ket{\mathrm{vac}}\,,
	\end{align}

	\item We also demand controllable cavity-assisted transitions that can be engineered to transfer excitations between levels $c$ and $b$~\cite{gupta07a,ritsch13a,landig16a}. These transitions induce a long-range interaction term, $J_c/N_M$, where we already include explicitly the inverse volume dependence of the cavity-assisted couplings. Besides, we still keep the local Raman assisted transitions between the $a$ and $b$ levels already used in section~\ref{subsec:spin}, with strength $g$ and detuning $\Delta$.
\end{itemize}

Summing up all these ingredients, the internal dynamics of the auxiliary atoms will be ruled by the following Hamiltonian:
\begin{equation}
\begin{split}
\label{eq:Haux3b}
\hat{H}\st{III,aux}&=\frac{J_c}{N_M}\sum_{\ii,\jj}\hat{B}_\ii^\dagger \hat{B}_\jj+ \Delta \sum_\jj \hat{B}_\jj^\dagger \hat{B}_\jj 
\\
&+J_A\sum_{\mean{\ii,\jj}}\hat{A}^\dagger_\ii\hat{A}_\jj+g\sum_\jj (\hat{A}_\jj^\dagger \hat B_\jj+\mathrm{H.c.})\,,
\end{split}
\end{equation}
where $J_A$ is the super-exchange coupling strengths, that can be tuned from positive to negative~\cite{duan03a,trotzky08a}, and that we will consider here to be $J_A>0$. Note that, apart from the first term describing cavity-assisted transitions, this Hamiltonian for the spin excitation is formally identical to the mediating Hamiltonian $H\st{II,aux}$ of Eq.~\eqref{eq:2levelHamiltonian}.

To show the scalability of the proposal, we study directly the case when $N_e$ fermions are present in the system with positions $\{\jj\}=\{\jj_1,\dots,\jj_{N_e}\}$. Inspired by the previous sections, we study the fermion interaction induced when only a single spin excitation is present in the system, which is initially symmetrically distributed among all fermionic positions:
\begin{equation}
\ket{\phi^{(0)}_{+}}=\frac{1}{\sqrt{N_e}}\sum_{\{\jj\}}\hat{B}^\dagger_{\jj}\ket{\mathrm{Mott}}\,.
\end{equation}

From Eq.~\eqref{eq:Haux3b}, it can be proven that $[\hat{H}\st{III,aux},\sum_{\ii}\left(\hat{B}^\dagger_\ii B_\ii +\hat{A}^\dagger_\ii A_\ii\right)]=0$, such that the number of spin excitations in this Hamiltonian is conserved, allowing us to work in the single excitation subspace of the Hamiltonian $\hat{H}\st{III,aux}$. Thus, all the possible wavefunctions are captured by the following ansatz:
\begin{align}
\ket{\phi\st{III,m}(\{\jj\})}_\mathrm{aux}=\sum_\kk\left(\phi^A_{m,\lla\jj}(\kk) \hat{A}_\kk^\dagger+\phi^B_{m,\lla\jj}(\kk) \hat{B}_\kk^\dagger\right) \ket{\mathrm{Mott}}\,.\label{eq:wave3}
\end{align}

Then, in order to obtain an analytical expression of the energy of the symmetric configuration $E\st{III,+}(\{\jj\})$ including the energy shift of the fermions, $\hat{H}_{f-\mathrm{aux}}(\{\jj\})+\hat{H}\st{III,aux}$, we apply perturbation theory using:
\begin{align}
\hat{H}_0=\Delta \sum_{\jj}\hat{B}_\jj^\dagger \hat{B}_\jj + U\sum_{\{\jj\}} \hat{B}_\jj^\dagger \hat{B}_\jj\,,
\end{align} 
as the unperturbed Hamiltonian. At this level, there is a degeneracy of the order of the number of fermions, that the cavity will break. Then, we include
\begin{align}
\hat{H}\st{cav} &=\frac{J_c}{N_M}\sum_{\ii,\jj}\hat{B}_\ii^\dagger \hat{B}_\jj\,,\label{eq:Hc}\\
\hat{H}_A &=J_A\sum_{\mean{\ii,\jj}}\hat{A}^\dagger_\ii\hat{A}_\jj+g\sum_\jj (\hat{A}_\jj^\dagger B_\jj+\mathrm{H.c.})\,,\label{eq:Ha}
\end{align}
as the two perturbations over it. Using perturbation theory, we find that the eigenenergy of the unperturbed state $\ket{\phi^{(0)}_{+}}$, with unperturbed energy $\hat{H}_0\ket{\phi^{(0)}_{+}}= E\st{III,B}^{(0)} \ket{\phi^{(0)}_{+}}=(U+\Delta)\ket{\phi^{(0)}_{+}}$, is perturbed to first order by the $\hat{H}_c$ leading to: $E\st{III,B}^{(1)}=U+\Delta+\rho_M J_c$, where  $\rho_M=N_e/N_M$ is the fermionic density, because the cavity breaks the degeneracy between the symmetric/antisymmetric wavefunctions, creating an energy difference $\rho_M J_c$ between them. In the next order, $\hat{H}_A$ leads to an additional correction of the energy which introduces the desired spatial dependence:

\begin{align}
E\st{III,+}^{(2)}(\{\jj\})&\approx E\st{III,B}^{(1)}  
+\frac{g^2}{N_e}\frac{1}{N_M}\sum_\kk \frac{\abs{e^{i\kk\cdot\jj_1}+\ldots+e^{i\kk\cdot\jj_{N_e}}}^2}{ E\st{III,B}^{(1)}-\omega_{\text{III},\kk}}\,,
\end{align}
where $\omega_{\text{III},\kk}=2 J_A\sum_{\alpha=x,y,z}\cos(k_\alpha)$ is the eigenenergy of the Hamiltonian $J_A\sum_{\mean{\ii,\jj}}\hat{A}^\dagger_\ii\hat{A}_\jj$ using periodic boundary conditions. In that equation, we observe that $\hat{H}_A$ delocalizes the auxiliary $A$-spin excitations providing the position dependent part of $E^{(2)}\st{III,+}(\{\jj\})$, which can be broken into a constant and a sum of pair-wise contributions which have the same shape as the one in Eq.~\eqref{eq:e2b}. In the continuum limit, $N_M\rightarrow \infty$, the pair-wise contributions can again be written as an integral that yields:
\begin{align}
E^{(2)}\st{III,+}(\{\jj\})&\approx E\st{III,B}^{(2)}+V\st{III}\sum_{\ii,\jj} \frac{\text{a}e^{-|\ii-\jj|/L\st{III}}}{|\ii-\jj|}\,,\label{eq:E3}
\end{align}
where, to second order, $E^{(2)}\st{III,B}= E\st{III,B}^{(1)} +\frac{g^2}{\mathcal{N}} \sum_\kk \frac{1}{E\st{III,B}^{(0,1)}-\omega_\kk} $. 
Here, $L\st{III}$ is the effective length of the Yukawa-Type potential, now given by, 
$L\st{III}=\text{a}\pa{E\st{III,B}^{(1)}/J_A-6}$, and $V\st{III}=\frac{g^2}{2\pi J_A N_e}$ its strength.

For self-consistency in the derivation of $E^{(2)}\st{III,+}(\{\jj\})$, in Eq.~\eqref{eq:E3}, we must impose that the corrections to the unperturbed state, $\ket{\phi^{(0)}_{+}}$, due to different elements of the Hamiltonian are small [Eq.~\eqref{eq:cond2}]. Deriving these contributions one by one,

\begin{figure}[tbp]
	\centering
	\includegraphics[width=0.7\linewidth]{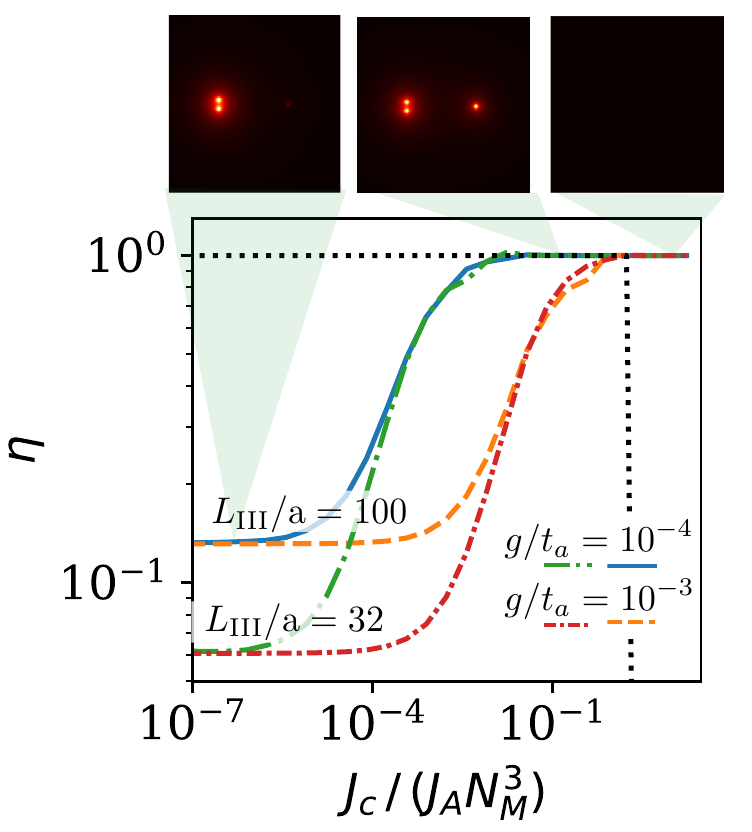}
	\caption{Exact diagonalization calculation of the bound state associated to Hamiltonian~\eqref{eq:Haux3b} for three fermions in the isosceles configuration illustrated in Fig.~\ref{fig:2cavity_no}. We represent the ratio $\eta$ between the population in bath $a$ of the fermion at the apex of the triangle, and one vertix of the base, as a function of the cavity strength. 
	The no-cavity limit ($J_c/J_A=0$) is determined by the fermionic geometry and effective length $L\st{III}/a$. 
	As the cavity interaction increases, inequality~\eqref{eq:wcond2} defines the lower cavity strength limit at which the population of the bath is equal for each of the fermionic positions.  The black dotted line shows the total population of atoms in level $A$ at the position of the three vertices ($W$ as defined in the main text), which is close to 1 for $J_c<U$, and quickly decays to an uniform distribution among all sites when the cavity interaction dominates the on-site interaction with the fermion $J_c>U$. 
	Inset: population of bath a for different values of cavity interaction. Note in the last inset that all the bath is equally populated when the cavity strength overpasses the on-site interaction $U$. 
	\textit{Parameters:} $U=2J_A$, $\Delta_c=10J_A$, $N_M=160^3$, $s/\text{a}=6$, $d/\text{a}=24$.}
	\label{fig:2cavity_comp}
\end{figure}

\begin{figure}[tbp]
	\centering
	\includegraphics[width=0.7\linewidth]{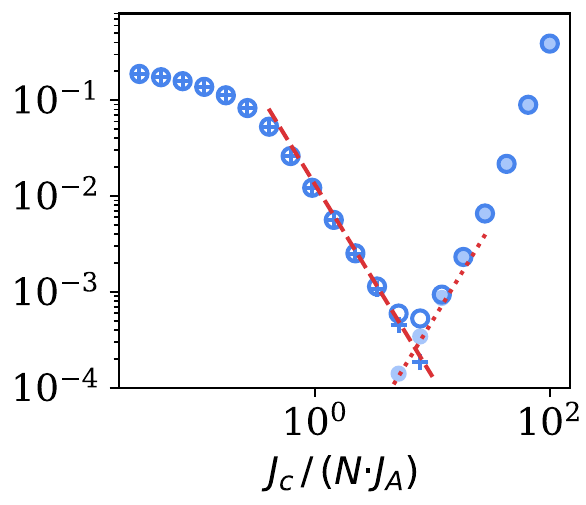}	
	\caption{Population of the mediating atom ground-state on contributions not corresponding to the leading-order ground-state, $\pa{\hat B_{\jj_1}+\hat B_{\jj_2}}\hat f_{\jj_1}\hat f_{\jj_2}/\sqrt{2}$, for any combination of $\jj_1, \jj_2$ (contoured marker), see main text. Here we use an exact diagonalization of the ground-state energy for a minimal model where two fermions are hopping on a 1D lattice, are attracted by two nuclei separated $8$ lattice sites, repel each other with an effective potential $V_0/d$ (being $d$ the interfermionic separation) and experience an on-site interaction $U$ with a bosonic species connected to a cavity mode (i.e. the terms associated to the fermionic dynamics and population in state $b$ of the Hamiltonian \eqref{eq:Haux3b}). Plus markers show the total population of the antisymmetric states of the form $\pa{\hat B_{\jj_1}-\hat B_{\jj_2}}\hat f_{\jj_1}\hat f_{\jj_2}/\sqrt{2}$ and coloured markers the population of sites not occupied by the fermions, $\hat B_{\rr}\hat f_{\jj_1}\hat f_{\jj_2}$ for $\rr \notin \lla{\jj_1, \jj_2}$. These are compared to the scaling predicted by the first-order analytical predictions given by Eq. \eqref{eq:wcond2} (red dashed line) and Eq. \eqref{eq:epsCav} (dotted line), respectively. Note that the dynamics in level $a$ is not included in this minimal model. \textit{Parameters:} $N=50^3,\,V_0=1/4J_A $, $U=4000J_A$.}
	\label{fig:Borcheck}
\end{figure}

\begin{itemize} 
	\item The cavity $\hat{H}_c$ Hamiltonian tends to delocalize the auxiliary atomic excitations beyond the fermion positions, which does not occur when the fermion-auxiliary atom interaction is large enough. Using Eq.~\eqref{eq:cond2}, we find that the cavity-mediated population of other symmetric states rather than $\ket{\phi_{+}^{(0)}}$ is upper bounded by:
    \begin{equation}
    \label{eq:epsCav}
        	\varepsilon\st{cav}=\abs{\frac{\sqrt{\rho_M}J_c}{U-J_c}}^2\,.
    \end{equation}
	such that one sufficient condition to satisfy $\varepsilon\st{cav}\ll 1$ is:
	\begin{equation}
	    J_c\ll U\,.\label{eq:wcond1}
	\end{equation}

	This is numerically confirmed in Fig.~\ref{fig:2cavity_comp}, where we study a three fermion configuration discussed in Fig.~\ref{fig:2cavity_no} using now the Hamiltonian $\hat{H}_{f-\mathrm{aux}}(\{\jj\})+\hat{H}\st{III,aux}$. For illustration, we plot the weight of the wavefunctions in the fermionic positions, i.e., $W=\sum_{\ii\in \{\jj\}}|\phi^A_{m,\lla\jj}(\ii)|^2$ (black dashed line in Fig.~\ref{fig:2cavity_comp}), as a function of $J_c/J_A$ for a fixed $U$ and for several values of $g/J$. There, we observe that $W\approx 1$ when Eq.~\eqref{eq:wcond1} is satisfied, irrespective of the particular choice of the rest of the parameters.
	
	\item As it occurred in subsection~\ref{subsec:spin}, the hoppings in $\hat{H}_A$ connect  $\ket{\phi_{+}^{(0)}}$ with two different set of states: (i) it dresses it with some population in the $A$-modes; and (ii) it takes it out of the symmetric sector. One can upper bound the corrections due to these two processes by $\varepsilon_A=\varepsilon_{A,i}+\varepsilon_{A,ii}$, where:
	\begin{equation}
	\label{eq:popAfirstOrder}
	\begin{split}
	\varepsilon_{A,i} &= \frac{g^2}{N_e N_M} \sum_\kk \abs{\frac{e^{i\kk\cdot\jj_1}+\ldots+e^{i\kk\cdot\jj_{N_e}}}{E\st{II,B}^{(0)}-\omega_{II,\kk}}}^2 \\
	& \leq \frac{g^2L\st{III}}{8\pi\text{a} J_A^2}N_e \ll 1\,,
	\end{split}
	\end{equation}
assuming the desired condition $d/L\st{III}\ll1$ for any pair of fermions, so that the Coulomb scaling prevails over the exponential decay. One observes that the final inequality scales as $g^2/J_A^2 \lesssim a/L\st{III}$, similarly to the two-fermion condition we encountered in the previous scheme (see Eq.~\eqref{eq:model2Aux}).
	
	The other contribution coming from the antisymmetric states is prevented by the energy gap between the symmetric/antisymmetric sector induced by the cavity-assisted transitions ($\rho_M J_c$), and it can be upper bounded by:
	\begin{align}
	\label{eq:modelIIIantis}
	\varepsilon_{A,ii}=\left(\frac{V\st{III
	}}{\rho_M J_c }\right)^2 G(\{\jj\})\,,
	\end{align}
	where $G(\{\jj\})$ is a function that solely depends on the particular fermionic configuration (see Appendix \ref{ap:perturbation}). Interestingly, $G(\{\jj\})\equiv 0$ in the case where all the fermions are equally spaced or when there are only two fermions, while in general it can always be upper-bounded by $\abs{G(\{\jj\})} \leq \pa{N_e/2-1}$. Then, the inequality to be satisfied when many fermions are present reads as:
	\begin{equation}
	        \varepsilon_{A,ii}\leq \left(\frac{V\st{III}}{\rho_M J_c }\right)^2 \frac{N_e}{2} \ll 1  \,.\label{eq:wcond2}
	\end{equation}
	
	This condition is also numerically benchmarked for the case triangular configuration of three fermions represented in Fig.~\ref{fig:2cavity_comp}. As in Fig.~\ref{fig:2cavity_no}, we plot the ratio of the weight of the wavefunction in the basis positions compared to the apex ($\eta$, see scheme in Fig. \ref{fig:2cavity_no}), showing how they only become equal in the limit when Eq.~\eqref{eq:wcond2} is satisfied.

  \item Besides, an extra condition appears to avoid that $\hat{H}_f$ connects the mediating state with the rest of the subspace (see Eqs.~\eqref{eq:cond2a}-\eqref{eq:cond2}). We can upper bound this contribution coming from the antysimmetric distribution of spin excitations at the fermionic positions by (see Appendix \ref{ap:perturbation}):
\begin{equation}
\label{eq:wcond4}
\begin{split}
&\varepsilon_{f} \approx \frac{\text{a}}{L\st{III}N_e}\pa{\frac{t_f}{\rho_M J_c}}^2\ll 1 \,.
\end{split}
\end{equation}

Testing this inequality numerically in a three-dimensional model is an outstanding challenge as it involves the three-dimensional Hilbert space of both the fermion and spin excitations in the $a$ and $b$ levels. Instead, in Fig.~\ref{fig:Borcheck}, we test Eq.~\eqref{eq:wcond4} in a minimal model of two fermions hopping in a 1D lattice for different values of the cavity coupling $J_c$. We observe a qualitative good agreement with the scaling $\propto (t_f/J_c)^2$ before the error introduced by an excessive cavity strength appears (Eq.~\eqref{eq:epsCav}).

\item Also as it occurred in the previous section, there is an additional condition to force that $L\st{III}$ does not vary depending on the fermionic configuration, as this will imply that the effective repulsive potential will change as the fermions hop to the lattice. Making an energy argument analogous to the derivation used in Eq.~\eqref{eq:lIIcondition}, one would desire $\abs{E\st{III,+}-E\st{III,B}^{(1)}}\ll (a/L\st{III})^2J_A$. This bound will highly depend on the particular fermionic configuration. An (unrealistic) upper bound for electronic repulsion would correspond to the case where all fermions are as close to each other as they can be, while respecting their fermionic character. In the limit of many simulated electrons, this scales as $V\st{III}N_e^{5/3}\ll (a/L\st{III})^2J_A$. This, however, does not make use of the entire allowed space for the fermions. A more realistic bound, taking into consideration that fermions will distribute in the entire lattice for the optimal simulation, one can approximate the repulsive energy as \footnote{See Eq. (6.6.19) of Ref.~\cite{Parr1989}} $\abs{E\st{III,+}-E\st{III,B}^{(1)}}\approx \frac{V\st{III}}{2^{1/3}}\pa{N_e-1}^{2/3}\sum_\jj \rho(\jj)^{4/3}$, where $\rho(\jj)$ denotes the fermionic density in site $\jj$. This estimation can then be particularized for any atomic/molecular level of interest. As a back-of-the-envelope calculation, considering an homogeneous density $\rho(\rr)\approx \rho_e$, one obtains,
\begin{align}
\label{eq:lengthcondition}
    \frac{V\st{III}}{2^{1/3}}\pa{N_e-1}^{2/3}N_e^{4/3}/N^{1/3} \ll (\text{a}/L\st{III})^2J_A\,.
\end{align}
The left-hand side of this estimation corresponds to the repulsion of an homogeneous distribution of $N_e$ atoms in a cubic lattice of $N$ sites. Intuitively, one can see that its scaling $\propto N_e^2/N^{1/3}$ corresponds to the previous unrealistic scaling $N_e^{5/3}$ of a cubic array of distance 1, corrected by the new characteristic length $\pa{N_e/N}^{1/3}$ when all the lattice is occupied~\footnote{Note that there is an errata in Eq. (10) of the printed version of Ref. \cite{arguello2019analogue}. There, an additional transformation $V_0\to V_0 \rho_e^{1/3}$ was assumed to account for the characteristic size of the molecule. This effect is already accounted without further assumptions when all inequalities are satisfied.}. 

\item Finally, there is an additional condition that only involves the localization length $L\st{III}$ of the Yukawa-type potential and the sizes of the fermionic/auxiliary atom lattice, that is, 
\begin{align}
\label{eq:wcond5}
N^{1/3}\ll L\st{III}/\text{a} \ll N_M^{1/3}\,,
\end{align}
whose intuition is clear: the length of the Yukawa potential has to be larger than then number of sites of the fermionic optical lattice ($N$), such that the fermions repel with a $1/r$-scaling, but smaller than the auxiliary atomic optical lattice, in order not to be distorted by finite size effects. In particular, one can relax condition \eqref{eq:lengthcondition}, aimed to ensure that the effective length in the Yukawa potential is constant regardless the fermionic configuration, and impose instead that the smallest and largest of them are contained within the range $\co{N^{1/3},N_M^{1/3}}$.

\end{itemize}

\section{\texorpdfstring{Benchmarking the simulator with two electron atoms (He) and molecules (HeH$^+$)}{Benchmarking the simulator with two electron atoms (He) and molecules (HeH+)}}
\label{sec:benchmark}
After having explained how to simulate all the elements of the quantum chemistry Hamiltonian in  a grid basis representation [Eqs.~\eqref{subeq:kin}-\eqref{subeq:rep}], here we illustrate the performance of our simulator for two-electron systems beyond the H$_2$ molecule considered in Ref.~\cite{arguello2019analogue}. In particular, we study in detail the simulation of the He atom (in subsection~\ref{subsec:helium}), that we will use to illustrate how to explore the physics of different spin symmetry sectors; and the HeH$^{+}$ molecule (in subsection \ref{subsec:molec}) to illustrate molecular physics for the case of unequal nuclei charges.

 For simplicity, we will use directly the $\hat{H}_\mathrm{eff}$ of Eq.~\eqref{eq:Heff} with a $V(\ii-\jj)=V_0 \text{a}/|\ii-\jj|$, assuming that all the simulator conditions are satisfied. Despite the apparent simplicity of the problem, obtaining the ground state energy of $\hat{H}_\mathrm{eff}$ for two fermions with, e.g., exact diagonalization methods, poses already an outstanding challenge since the number of single-particle states in a grid basis scale with the number of fermionic lattice sites $N$. To obtain the results that we will show in the next subsections, we have then adopted an approach reminiscent of Hartree Fock methods, where the Hamiltonian is projected in a basis that combines atomic states calculated from the single-particle problem, together with electronic orbitals that interact with an average charge caused by the rest of electrons (see Appendix \ref{ap:nummethods} for more details). Let us remark that these are just numerical limitations to benchmark the simulator, that would be free from these calculations details. 

\subsection{He atom}
\label{subsec:helium}
The first system we consider is the case of the He atom. This corresponds to a system with a single nucleus with $Z_1=2$, such that one just requires a single spatially-shaped laser beam to mimic the nuclear potential, and two simulated electrons. Let us note that since $\hat{H}_\mathrm{eff}$ does not couple the position and spin degree of freedom, one can solve independently the problems where the spin degrees of freedom are in a singlet (antisymmetric) or triplet states (symmetric), that will result in spatially symmetric/antisymmetric wavefunctions, traditionally labeled as para- and orthohelium, respectively.

\begin{figure}[tb]
	\centering
	\includegraphics[width=\linewidth]{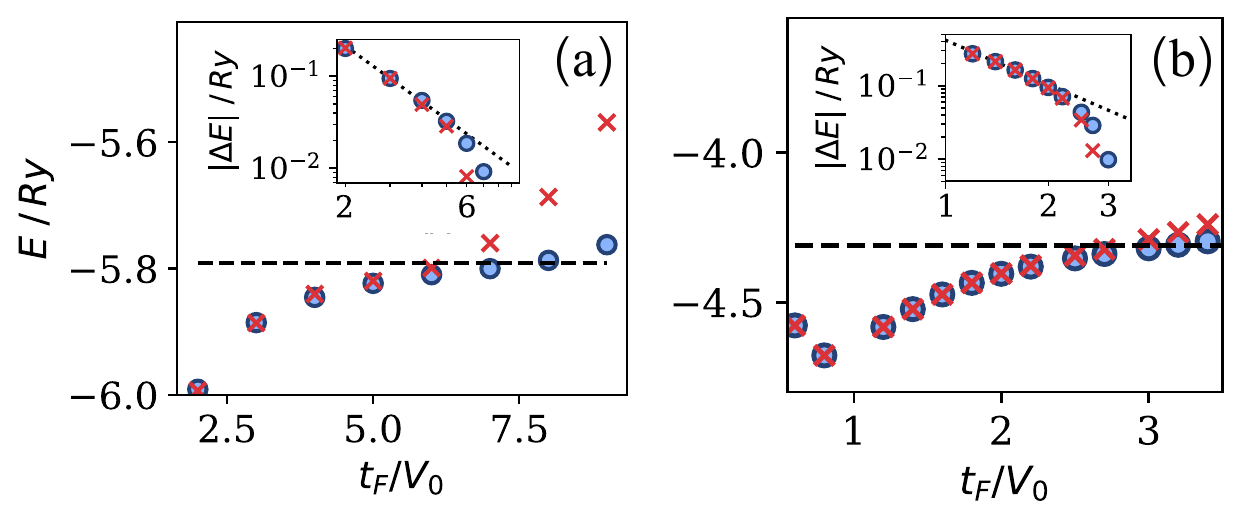}
	\caption{Ground state energy energy of the discretized Hamiltonian associated to atomic He, in the ortho (a), and parahelium sectors (b). Following the extrapolation method, dashed lines indicate the values $E_{\mathrm{para,He}}^\infty =-5.79$ Ry and $E_{\mathrm{ortho,He}}^\infty =-4.31$ Ry, respectively, for which the scaling of the energy error as $(\text{a}_0/\text{a})^{-2}$ is observed (insets). Round (crossed) markers correspond to $N=100^3$ ($N=75^3$).}
	\label{fig:atHe}
\end{figure}

In Fig.~\ref{fig:atHe}(a) we plot the ground state energy of the He atom as a function of the effective Bohr radius presented in Eq.~\eqref{eq:Bohr}. Note that there is no closed solution already for this very simple system, and our simulator will be compared to numerical results with no relativistic or QED corrections~\cite{pekeris1959}. 
Furthermore, we use the extrapolation strategy explained in Section~\ref{subsec:errors} based on the scaling of the error $\Delta E/Ry\propto (a_0/d)^{-2}$ to obtain the expected energy that will come out from the simulation, yielding $E_{\mathrm{para,He}}^\infty =-5.79$ Ry, and $E_{\mathrm{ortho,He}}^\infty =-4.31$ Ry. Their relative error to the respected tabulated values~\cite{pekeris1959}, $-4.3504$ Ry and $-5.8074$ Ry, is therefore of $0.3\%$ and $0.9\%$ respectively for the benchmarking done with a system $N=100^3$. Note that the bigger error corresponds to orthohelium, whose orbitals are larger and thus more affected by the discretization of the lattice. 

\subsection{\texorpdfstring{HeH$^+$ molecule}{HeH+ molecule}}
\label{subsec:molec}
Here, we study the two-electron molecule  He$^{+}$-H, which has two nuclei, one with charge $Z_1=1$ (the one corresponding to the H atom) and another one with $Z_2=2$ (the one corresponding to the He cation). Thus, the simulator requires two spatially-shaped laser beams, one with double the intensity than the other, such that its induced potential is twice as big.

One of the magnitudes of interest in molecular physics is the molecular potential, that is, how the ground state energy of the molecule varies as a function of the distance $d$ between the nuclei. This curve already provides useful information, such as its equilibrium molecular position (if any) as well as its dissociation energy. In our simulator, in order to always maintain the nuclei half a site away from the nodes of the lattice (to avoid a divergent value of the potential) we choose integer values of $d/\text{a}$.

\begin{figure}[tb]
	\centering
	\includegraphics[width=0.9\linewidth]{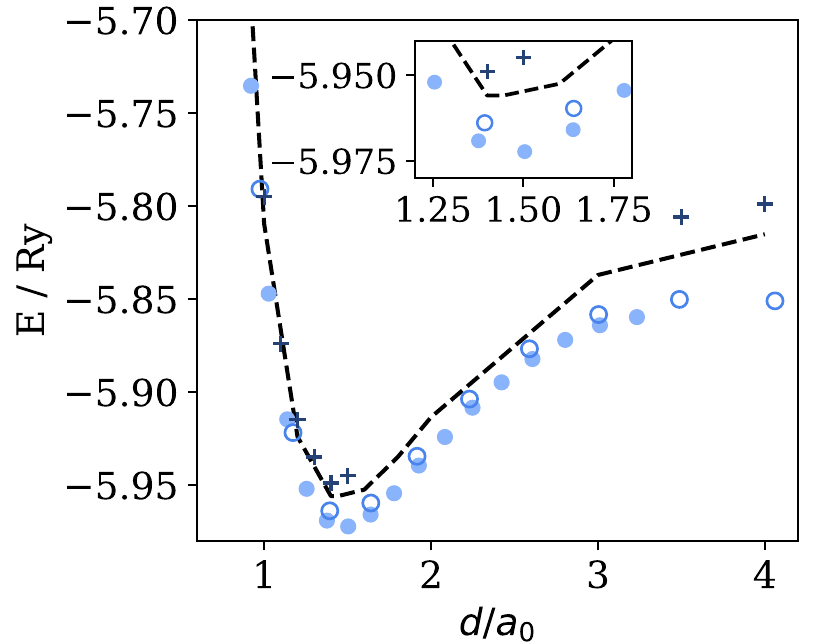}
	\caption{Molecular potential of $\text{HeH}^+$ as a function of the lattice size calculated for a finite lattice of $N=75^3$ (coloured markers) and $N=100^3$ (contoured markers). Crossed markers follow the mitigation strategy using these sizes, and described in Sec.~\ref{subsec:errors}. Black dashed line follows the molecular potential beyond discretization numerically calculated~\cite{Kolos1985}. 
	Inset zooms around the position of the minimum.}
	\label{fig:molHeH}
\end{figure}

In Fig.~\ref{fig:molHeH} we plot the molecular potential that could be obtained with our simulator for two different system sizes $N=75^3$ and $N=100^3$. As we did before, for each value of $d/\text{a}_0$, we choose the optimal discrete Bohr radius, $\text{a}_0/\text{a}$, using the extrapolation strategy explained in section~\ref{subsec:errors}. Notice that this molecular potential needs to include nuclear repulsion, and its minimum corresponds to the distance at equilibrium. The energy at this point, $E_{\mathrm{min,HeH^+}}^\infty =-5.95$ Ry, which is in agreement with the numerical value $-5.95740408$ Ry reported in~\cite{Kolos1985}. As the separation increases, we observe that the error of the finite simulator increases, since the finiteness of the lattice is more restrictive when more sites need to separate the nuclear position. We also observe that the continuum result obtained with the mitigation approach still tends to the dissociation limit corresponding to ortho-Helium, discussed in the previous sections.

\section{Conclusion and outlook}
\label{sec:conclusion}

Summing up, we have expanded the analysis of the original proposal of Ref.~\cite{arguello2019analogue} on how to simulate quantum chemistry Hamiltonians in an analog fashion using ultra-cold fermionic atoms in optical lattices. In particular, this work provides several original results, such as: i) A discussion of the physics of the holographic potentials required to obtain the nuclear attraction term. ii) The introduction of two simplified setups to obtain fermionic repulsion. Although the emergent interactions are not fully Coulomb-like, these simpler setups can already be used as intermediate, but meaningful, experiments to observe chemistry-like behaviour~\cite{Arguello-Luengo2020b}, and to benchmark existing numerical algorithms. iii) An extrapolation strategy which allows us to obtain the expected energies in the continuum limit beyond the limitations imposed by the finite size of the simulator and, importantly, without an a priori knowledge of the expected energy. This approach could also guide other systems simulating chemistry problems in a lattice. iv) A numerical benchmark of the working conditions of the simulator. v) Finally, an illustration of the simulator capabilities for two-electron systems like the He atom and the HeH$^+$ molecule.

Taking this work as basis, there are many interesting directions that one can pursue. A particularly appealing one in the near-term is to continue simplifying the ingredients required for the proposal, even at the cost of not simulating real chemistry~\cite{Arguello-Luengo2020b}. Another one would be the study of dynamical processes, e.g., chemical reactions or photo-assisted chemistry, that is typically very hard numerically, and where the slower timescales of our simulator and the excellent imaging techniques can provide real-time access to the wavefunction properties. Finally, given our ability to tune the effective fermion interaction, one can use a different bound state to mediate attractive interactions. This would allow us to simulate chemistry beyond Born-Oppenheimer approximation by including another atomic specie that plays the role of the nuclei. Beyond the chemistry simulation, we also envision that the method to engineer non-local interactions in ultra-cold atoms can be exported to explore other phenomena where that type of interactions play a role, e.g., like in long-range enhanced topological superconductors~\cite{viyuela18a}.

\section*{Acknowledgements}
The authors acknowledge very insightful discussions and feedback from J. I. Cirac and P. Zoller, with whom they worked in the original proposal of Ref.~\cite{arguello2019analogue}.
J.A.-L. acknowledges support from 'la Caixa' Foundation (ID 100010434) through the fellowship LCF/BQ/ES18/11670016, the Spanish Ministry of Economy and Competitiveness through the 'Severo Ochoa' program (CEX2019-000910-S), Fundaci{\'o} Privada Cellex, Fundaci{\'o} Mir-Puig, and Generalitat de Catalunya through the CERCA program and QuantumCat (001-P-001644). 
T. S. acknowledges the support from NSFC 11974363.
A. G.-T. acknowledges support from the Spanish project PGC2018-094792-B-100 (MCIU/AEI/FEDER, EU) and from the CSIC Research Platform on Quantum Technologies PTI-001.

\setcounter{equation}{0}
\setcounter{figure}{0}
\setcounter{table}{0}
\setcounter{section}{0}
\makeatletter

\renewcommand{\thefigure}{A\arabic{figure}}
\renewcommand{\thesection}{A\arabic{section}}  
\renewcommand{\theequation}{A\arabic{equation}}  

\newpage
 \widetext
\section{Discretization error scaling}
\label{ap:discEffects}
This discretization inherent to our lattice approach gives rise to certain errors that need to be considered and vanish in the infinite-size limit. Here we estimate the errors related to (i) the disretization of the integral appearing int the Coulomb term, and (ii) the discretization of the Laplacian appearing in the kinetic term. 

\subsection*{Discretization of the integrals.} 
The calculation of expected energies over the continuum are based on integrals on the entire real space. The discretization of the lattice, however, transforms these integrals into a finite sum of terms, introducing an error that vanishes in the limit of infinite sites. This effect is closely related to the definition of a Riemann integral evaluated at mid-point values, $\mm_\jj=\jj+(0.5,0.5,0.5)$, written in the units of the lattice spacing. Its error is given by,
\begin{align}
\Delta_\textrm{Int} & =\abs{\int\ d\sss f(\sss)-\text{a}^3\sum_\jj f(\mm_\jj)} \approx \frac{\text{a}^2}{24}\int d\sss \co{f_{xx}(\sss)+f_{yy}(\sss)+f_{zz}(\sss)}\,,
\end{align} 
where $f_{\alpha\beta}(\xx)=\partial_\alpha \partial_\beta f(\xx)$. A back-of-the-envelope calculation could be illustrative in this case. We focus on the integrals for the Coulomb potential, $f(\rr)=|\psi_{n\ell}(\rr)|^2\cdot V(\rr)$, for Hydrogen atomic orbitals, $\psi_{n\ell}(\rr) \propto g_{n\ell}(r/a_0)\cdot \exp(-nr/\text{a}_0)$; being $g_{n\ell}$ an ($n-1$)-degree polynomial. Rescaling coordinates the lattice units, $r\to r\text{a}_0$, one has
$$
\Delta ^{(C)}_\textrm{Int} \equiv V_0 \frac{a^3\int d\rr\; \frac{\partial^2}{\partial x^2}\co{g_{n\ell}^2(r)\cdot \exp(-2nr)/r}}{a_0^3\int d\rr\; g_{n\ell}^2(r)\cdot \exp(-2nr)}\propto V_0\pa{\text{a}/\text{a}_0}^3\,.
$$
Expressing this in Rydberg units, one gets the scaling, 
\begin{equation}
\label{eq:coulError}
\Delta ^{(C)}_\textrm{Int}/Ry\propto \pa{V_0/t_F}^2\,.
\end{equation}
The precise constant accompanying this scaling is a geometrical factor, characteristic of each atomic orbital. 

\subsection*{Approximation of the kinetic term.} 
In the discrete Hamiltonian, the kinetic term is approximated as a first-neighbor hopping term. One can estimate the error in this approximation from the next order terms of the expansion of $\nabla^2 f$, that correspond to $\frac{a^4}{12}\co{\partial_x^4 f(\rr)+\partial_y^4 f(\rr)+\partial_z^4 f(\rr)}$. 

Again, one can make an estimation on how this error of the kinetic term scales with the atomic units, and therefore with the size of the system. Using the Hydrogen wavefunctions used before, one gets, 
$$
\Delta ^{(L)}_\textrm{lin} \equiv t_F \frac{a^4\int d\sss \;  g_{n\ell}(r)\exp(-r/n)\cdot  \frac{\partial^4}{\partial x^4}\co{g_{n\ell}(r) \exp(-nr)}}{a_0^4\int d\rr\; g_{n\ell}^2(r)\cdot \exp(-2nr)}\propto t_F(\text{a}/\text{a}_0)^4
\,.$$
Expressing this result in atomic units, one obtains the leading correction for the final error in energies,   
\begin{equation}
\label{eq:kinError}
\Delta ^{(L)}_\textrm{lin}/Ry\propto \pa{V_0/t_F}^2\,.
\end{equation}

Interestingly, both effects lead to the heuristic scaling $\propto \pa{V_0/t_F}^2$ for discretization error that we numerically observe in Fig. \ref{fig:hydrQual}.

\section{Details on the full perturbation theory analysis of Section~\ref{sec:qcelectron}.}
\label{ap:perturbation}

Here, we complete the details on the derivation of the bounds presented in the main text. In our derivation, one is interested in finding the mediating species in the state providing repulsion. Following the approach introduced in Eq.~\eqref{eq:cond2a}, our bound will arise from the coupling of this state to other orthogonal ones, and the energy gap between them, $ \varepsilon_\alpha=\sum_{\mathrm{all}\{\rr\}}\sum_{m}\bar\varepsilon_\alpha\pa{\varphi^\perp_{m,\{\rr\}}}$, with 

\begin{align}
    \bar\varepsilon_\alpha\pa{\varphi^\perp_{m,\{\rr\}}}=\left|\frac{_f\bra{\{\jj\}}_\mathrm{aux}\bra{\varphi^\perp_{m,\{\rr\}}}\hat{H}_\alpha\ket{\Psi}}{\Delta_{m,\rr,\jj}}\right|^2\,.
\end{align}

\subsection*{Useful analytical expressions}

Here, we first derive the analytical expressions of certain integrals that appear several times in the calculations of the error bounds. These are expressions of the form:
\begin{equation}
\Sigma(z,\rr)=\frac{1}{(2\pi)^3}\int_D d\kk \frac{e^{i\kk\cdot\rr}}{z-\omega(\kk)}\,,
\end{equation}
for $D=\co{-\pi,\pi}^{\otimes 3}$, and $\omega(\kk)=2t\co{\cos(k_x)+\cos(k_y)+\cos(k_z)}$ (we assume $t\equiv 1$ from now on), that, for example, governs the shape of the single-fermion bound-state wavefunction. Other expressions that appear are of the type:
\begin{equation}
g(z,\rr)=\frac{1}{(2\pi)^3}\int_D d\kk \frac{e^{i\kk\cdot\rr}}{\co{z-\omega(\kk)}^2}\,,
\end{equation}
that governs the Franck-Condon coefficient in the same situation. Note that the latter is related to $\Sigma(z,\rr)$ by a derivative: $g(z,\rr)=-\partial_z \Sigma(z,\rr)$. Remarkably, in the limit $\kk\cdot\rr \gg 1$, one can expand the dispersion relation around their band-edges, $\omega(\kk)\approx 6-\kk^2$, and extend the integration domains to infinite to obtain an analytical expression:
\begin{equation}
\label{eq:apsigr}
\begin{split}
\Sigma(z,\rr)&=\frac{1}{(2\pi)^2} \int_0^{\pi} d\theta \int_0^\infty  dk \, \frac{e^{ik\cdot r\cos\theta}}{(z-6)+k^2}\, k^2\sin\theta=\frac{1}{(2\pi)^2} \int_{-1}^{1} ds \int_0^\infty  dk \, \frac{e^{ik\cdot rs}}{(z-6)+k^2}\, k^2 \\
&=\frac{-i}{r(2\pi)^2} \int_0^\infty  dk \, \frac{e^{ik\cdot r}-e^{-ik\cdot r}}{(z-6)+k^2}\, k=
\frac{-i}{r(2\pi)^2} \int_{-\infty}^\infty  dk \, \frac{e^{ik\cdot r}}{(z-6)+k^2} \\  
&= \frac{1}{4\pi r}e^{-r\sqrt{z-6 }} \,.
\end{split}
\end{equation}

Note that for $\rr=[0,0,0]$ the integral does not converge, because we have artificially introduced a divergence by expanding the domain of integration to infinite. A way of renormalizing consists in artificially introducing an exponential cut-off $e^{-k\Lambda}$ with $\Lambda \to 0$, such that:
\begin{equation}
\label{eq:apsig0}
\begin{split}
\Sigma(z,0)&=\frac{1}{2\pi^2} \int_0^\infty  dk \, \frac{k^2 e^{-k\Lambda}}{(z-6)+k^2} = \frac{1}{2\pi^2\Lambda}-\frac{\sqrt{z-6}}{4\pi}\,.
\end{split}
\end{equation}
However, in lattice systems this cutoff appears naturally, and one can analytically obtain an expression for $\Sigma(z,0)$~\cite{Joyce2002,Borwein1992}:
\begin{equation}
\label{eq:apsigz0}
\begin{split}
\Sigma(z,0)\approx 0.253-\frac{\sqrt{z-6}}{4\pi} \,.
\end{split}
\end{equation}

Once we have the analytical expansions of $\Sigma(z,\rr)$ it is straightforward to obtain the higher order terms, e.g., $g(z,\rr)$ as follows:
\begin{equation}
\label{eq:gzr}
        g(z,r)=-\partial_z \Sigma(z,\rr)=\frac{1}{8\pi \sqrt{z-6}} e^{-r\sqrt{z-6} }\,.
\end{equation}

\subsection*{Scheme I: Repulsion mediated by single atoms}
In the single-fermion case of Scheme I, the maximum ratio between the hopping of the fermionic and mediating species was obtained in Eq.~\eqref{eq:model1tftb} from the coupling to the scattering states of the mediating atom when the fermion hops: 
$$
\sum_m \bar\varepsilon_f\pa{\varphi^\perp_{m,\jj_0+1}} \leq  t_f^2 \sum_\kk \abs{\frac{ \hat b_\kk \ket{\varphi_{B,\jj_0}}}{E\st{I,B}-\omega_\kk}}^2$$ \,.

Note that in the three-dimensional lattice, the sum to nearest neighbors introduces a factor $6$, that we have omitted along the text to focus on the scalings. Without loss of generality, we can consider $\jj_0$ to be the origin of coordinates. Replacing the wavefunction \eqref{eq:1boundwave} in momentum-space, one obtains $\abs{ \hat b_\kk \ket{\varphi_{B,\jj_0}}}^2=1/\co{\mathcal{N}_B N_M\pa{E\st{I,B}-\omega_\kk}^2}$. 
Replacing the expression \eqref{eq:gzr}, and its second derivative, one obtains the inequality \eqref{eq:model1tftb}.

As we see, moving to momentum space simplifies the calculation of $\mathcal{F}_1$ in Eq.~\eqref{eq:f1},
\begin{align}
\braket{\varphi_{B,\jj_0+1}}{\varphi_{B,\jj_0}}=\frac{1}{\mathcal{N}_B N_M}\sum_\kk \frac{e^{ik_z}}{\pa{E\st{I,B}-\omega_\kk}^2}=e^{-a/L\st{I}}\,,
\end{align}
where we have made use of Eq.~\eqref{eq:gzr}.

Moving now to the two-fermion case, one can relate Eqs.~\eqref{eq:1boundstate} and~\eqref{eq:2boundState} to obtain,
\begin{equation}
    \frac{1}{N_M} \sum_\kk \frac{1}{E\st{I,B}-\omega_\kk}=\frac{1}{N_M}\sum_{\kk}\frac{1 + e^{i\kk\cdot\jj_{12}}}{E_+(\{\jj\})-\omega_{\kk}}\,,
\end{equation}
and then replacing the expression \eqref{eq:apsigr} and \eqref{eq:apsig0} leads to:
\begin{equation}
\label{eq:relationBS}
    -\sqrt{E\st{I,B}/t_b-6}=-\sqrt{E_+(|\jj_{12}|)/t_b-6}
    +\frac{\text{a}e^{-|\jj_{12}|\sqrt{E_+(d)/t_b-6}}}{|\jj_{12}|}\,.
\end{equation}

Here, we need to separate the discussion in two different regimes. 
\begin{itemize}
  \item In the case $|\jj_{12}|\gg L\st{I}/a$ the latter term in \eqref{eq:relationBS} is dominated by the exponential decay, and one can expand to lowest order the effective repulsive potential $V_{\text{I},>}(\jj_{12})$ in Eq. \eqref{eq:short1} by replacing $E_+(d)\approx E\st{I,B}$ in the exponential.

\item In the regime $|\jj_{12}|\ll L\st{I}/a$ that simplification is, however, not possible. A general expansion of Eq.~\eqref{eq:relationBS} in this regime corresponds to,
\begin{align}
    V_{\text{I},<}(\jj_{12})=\frac{\gamma^2\text{a}^2}{|\jj_{12}|^2} + \frac{2\gamma}{1+\gamma}\frac{\text{a}^2}{|\jj_{12}|L\st{I}}+\frac{\text{a}^2\mathcal O \co{\pa{|\jj_{12}|/L\st{I}}^2}}{|\jj_{12}|^2} \,,
\end{align}
with $\gamma\approx 0.567$. This leads to Eq.~\eqref{eq:long1}. 
\end{itemize}

In this exponential regime, we now need to bound the undesirable coupling to the antisymmetric state due to the fermionic hopping, $\bar\varepsilon_f\pa{\varphi_-\pa{\jj_1+1,\jj_2}}$. From the bound-state wavefunction of Eq.~\eqref{eq:wavefunction2}, neglecting terms exponentially suppressed by the distance, we obtain:
$$
\braket{\varphi_{-,\pa{\jj_1+1,\jj_2}}}{\varphi_{+,\pa{\jj_1,\jj_2}}}\approx \frac{1}{2\mathcal N_B N_M}\sum_\kk \frac{1-e^{ik_z}}{\pa{E\st{I,B}-\omega_\kk}^2}\approx \frac{1-\mathcal F_1}{2}\,.
$$

Within this regime, we can replace in the denominator $E_+\pa{\jj_1,\jj_2}-E_-\pa{\jj_1+1,\jj_2}\approx 2V\st{I}\pa{|\jj_{12}|}$, which leads to the result stated in Eq.~\eqref{eq:errorM1b}.

\subsection*{ Scheme II: Repulsion mediated by atoms subject to state-dependent potentials}

As detailed in the main text, in this scheme to gain tunability we included a second level in the mediating species. To analyze its effect, we can separate the total Hamiltonian into the unperturbed $(\hat H_0)$ and perturbed $(\hat H_1)$ terms:
\begin{align}
\begin{split}
\hat H_0 =&\Delta \sum_\jj  \hat{b}_\jj^\dagger \hat{b}_\jj+ U \sum_{\jj_i\in\{\jj\}}\hat{b}_{\jj_i}^\dagger \hat{b}_{\jj_i}  -t_a \sum_{\langle \ii , \jj\rangle }\hat{a}_\ii^\dagger \hat{a}_\jj-t_b \sum_{\langle \ii , \jj\rangle }\hat{b}_\ii^\dagger \hat{b}_\jj \,,
\\
\hat H_1 =&g\sum_\jj (\hat{b}_\jj^\dagger \hat{a}_\jj +\hc)
\,.
\end{split}
\end{align}

To lowest order, and assuming that $t_b$ is negligible, the ground state of $\hat H_0$ corresponds to $\ket{\varphi\st{II,+}^{(0)}}=(\hat b_{\jj_1}\dg+\hat b_{\jj_2}\dg)/\sqrt{2}\vac$, with energy $E\st{II,+}^{(0)}=U+\Delta$. The effective repulsion enters then as a second-order contribution in perturbation theory, 
\begin{align}
    E\st{II,+}^{(2)}=\frac{\abs{\braket{\text{vac}|\hat a_\kk \hat H_1}{\varphi\st{II,+}^{(0)}}}^2}{E\st{II,+}^{(0)}-\omega_{\text{II},\kk}}=\frac{g^2}{2N_M}\sum_{\kk}\frac{\abs{e^{i\kk\cdot\jj_1}+e^{i\kk\cdot\jj_2}}^2}{E\st{II,+}^{(0)}-\omega_{\text{II},\kk}}=
    \frac{g^2}{N_M}\sum_{\kk}\frac{1}{E\st{II,+}^{(0)}-\omega_{\text{II},\kk}}+\frac{g^2}{N_M}\sum_{\kk}\frac{e^{i\kk(\jj_1-\jj_2)}}{E\st{II,+}^{(0)}-\omega_{\text{II},\kk}}\,,
\end{align}
which corresponds to Eq.~\eqref{eq:e2b}.

To higher order, $\ket{\varphi\st{II,+}}$ will also have contribution in level $a$ of the form, $\ket{\varphi\st{II,+}}=\ket{\varphi\st{II,+}^{(0)}} +\alpha_\kk \hat a_\kk\vac$. To make this perturbative expansion valid, Eq.~\eqref{eq:model2Aux} bounds the first-order contributions as follows:
\begin{align}
\label{eq:firstOrdera}
  \varepsilon_\mathrm{aux}=\sum_\kk \abs{\alpha_\kk^{(0)}}^2= \sum_\kk \abs{\frac{\braket{\text{vac}|\hat a_\kk \hat H_1}{\varphi\st{II,+}^{(0)}}}{E\st{II,+}^{(0)}-\omega_{\text{II},\kk}}}^2 \approx \frac{g^2}{2 N_M}\sum_\kk \abs{\frac{e^{i\kk\cdot\jj_1}+e^{i\kk\cdot\jj_2}}{E\st{II,+}^{(0)}-\omega_{\text{II}\kk}}}^2\leq \frac{2g^2}{N_M}\sum_\kk \frac{1}{\pa{E\st{II,+}^{(0)}-\omega_{\text{II}\kk}}^2}=\frac{V\st{II} L\st{II}}{t_a \text{a}}\,.
\end{align}

While non-dominant, this population is relevant as it is responsible for the induced repulsion. To make the effective length independent on the particular fermionic configuration, in the derivation of Eq.~\eqref{eq:lIIcondition}, we have bounded the next-order contribution to this population. Exploring the next non-negligible order, we obtain:
\begin{align}
\label{eq:secondOrdera}
\sum_\kk &\abs{\alpha_\kk^{(0)}+\alpha_\kk^{(3)}}^2=\sum_\kk \abs{ \frac{\braket{\text{vac}|\hat a_\kk \hat H_1}{\varphi\st{II,+}^{(0)}}}{E\st{II,+}^{(0)}-\omega_{\text{II},\kk}}+\sum_\qq \frac{\abs{\braket{\text{vac}|\hat a_\qq \hat H_1}{\varphi\st{II,+}^{(0)}}}^2\braket{\text{vac}|\hat a_\kk \hat H_1}{\varphi\st{II,+}^{(0)}}}{\pa{E\st{II,+}^{(0)}-\omega_{\text{II},\kk}}\pa{E\st{II,+}^{(0)}-\omega_{\text{II},\qq}}}\pa{\frac{1}{E\st{II,+}^{(0)}-\omega_{\text{II},\kk}}+\frac{1}{2\pa{E\st{II,+}^{(0)}-\omega_{\text{II},\qq}}}}}^2 \\
\approx & \varepsilon_{aux}+
\sum_\kk \frac{\abs{\braket{\text{vac}|\hat a_\kk \hat H_1}{\varphi\st{II,+}^{(0)}}}^2}{\pa{E\st{II,+}^{(0)}-\omega_{\text{II},\kk}}^3} 
\sum_\qq \frac{\abs{\braket{\text{vac}|\hat a_\qq \hat H_1}{\varphi\st{II,+}^{(0)}}}^2}{\pa{E\st{II,+}^{(0)}-\omega_{\text{II},\qq}}}+\ldots \approx \frac{V\st{II} L\st{II}}{t_a \text{a}} + \frac{V\st{II} L\st{II}}{t_a \text{a}} \pa{\frac{g L\st{II}}{4t_a \text{a}}}^2+\ldots\,,
\end{align}
where, in the right hand side, we have applied Eqs.~\eqref{eq:apsigz0} and~\eqref{eq:gzr}; and omitted higher order terms in $\pa{\frac{g L\st{II}}{4t_a \text{a}}}^2$. In Fig.~\ref{fig:32checkCondition} we observed that imposing now that the second term in Eq.~\eqref{eq:secondOrdera} is smaller than the first one, e.g., using a ratio $\pa{\frac{g L\st{II}}{4t_a \text{a}}}^2=0.01$ (dotted line), allows for a constant definition of $L\st{II}$ for any fermionic configuration.

\subsection*{Scheme III: Repulsion mediated by atomic spin excitations and cavity assisted transitions}

In this final scheme, we include a cavity interaction to ensure a pairwise effective repulsion when more than two fermions are included in the system. There are several errors one should account for:

\begin{itemize}
\item The cavity also couples the unperturbed state $\ket{\phi^{(0)}_{+}}=\frac{1}{\sqrt{N_e}}\sum_{\{\jj\}}\hat{B}^\dagger_{\jj}\ket{\mathrm{Mott}}$, to the other symmetric state in positions not occupied by the fermions, $\frac{1}{\sqrt{N_M-N_e}}\sum_{\rr\notin\{\jj\}}\hat{B}^\dagger_{\rr}\ket{\mathrm{Mott}}$. The coupling between them has intensity $J_c \sqrt{N_e\pa{N_M-N_e}}/{N_M}$, and the energy difference is $U-\pa{1-2N_e/N_M}J_c$. Therefore, the error of Eq.~\eqref{eq:epsCav} one needs to bound is,
\begin{align}
    \varepsilon\st{cav}=\abs{\frac{J_c\sqrt{N_e\pa{N_M-N_e}}/{N_M}}{U-\pa{1-2N_e/N_M}J_c}}^2 \approx \abs{\frac{J_c\sqrt{N_e/N_M}}{U-J_c}}^2 
\end{align}
for $N_e\ll N_M$. 

\item Even if the cavity does not couple this state with other antisymmetric ones, this can still occur as a consequence of coupling $\hat H_A$. For $J_c\ll U$, the relevant energy gap corresponds to $U$, which separates the excitation of state $B$ in atoms placed at fermionic position, against unoccupied positions. Therefore, now we focus on the $N_e-1$ orthogonal states, that are also orthogonal to $\ket{\phi^{(0)}_{+}}$, and can be written as $ \ket{\phi^{(0)}_{\bot,m}}=\sum_{\rr\in\{\jj\}}\lambda_{m,\rr}\hat{B}^\dagger_{\rr}\ket{\mathrm{Mott}}$, with energy $U+\Delta$, and satisfying $\sum_{\rr\in\lla{\jj}}{\lambda_{m,\rr}}=0$ and $\sum_{\rr\in\lla{\jj}}\abs{\lambda_{m,\rr}}^2=1$ for every $m=1\ldots N_e-1$. The error due to the coupling to these states reads as,
\begin{align}
\varepsilon_{A,ii}=\sum_m & \bar\varepsilon_A\pa{\ket{\phi^{(0)}_{\bot,m}}}=\sum_m 
\abs{\sum_{\kk}  \frac{\bkev{\phi^{(0)}_{\bot,m}}{\hat H_A \hat A^\dagger_{\kk}}{\mathrm{Mott}}\bkev{\mathrm{Mott}}{\hat A_{\kk} \hat{H}_A}{\phi^{(0)}_{+}}}{\pa{E\st{III,B}^{(1)}-\omega_{\text{III},\kk}}\pa{E\st{III,B}^{(1)}-U-\Delta}}}^2 \nonumber \\
& = \pa{\frac{g^2}{\rho_M J_c}}^2 \frac{1}{N_e}\, \sum_{m}\abs{\frac{1}{N_M}\sum_\kk \frac{f_\kk (m)}{E\st{III,B}^{(1)}-\omega_{\text{III},\kk}}}^2  \,.
\end{align}
where $f_\kk (m)=\sum_{\sss,\rr \in \lla{\sss}; \rr\neq\sss} \lambda_{m,\sss} e^{i\kk (\jj-\rr)}$ accounts for the relative distances weighted by the components of the states involved. To upper-bound this sum, it translates after integration into,
$$
 \sum_{m}\abs{\frac{1}{N_M}\sum_\kk \frac{f_\kk (m)}{E_s-\omega_\kk}}^2
 \approx \sum_{m} \abs{\sum_{\sss \in \lla{\jj}} \sum_{\rr \in \lla{\jj}; \rr\neq\sss}  \lambda_{m,\sss} x_{\sss\rr}}^2\,,
$$
where $x_{\sss \rr}=\frac{1}{4\pi J_A}\frac{\text{a}}{\abs{\sss-\rr}}\in \frac{1}{4\pi J_A} (0,1]$. To give a base-independent argument, one can simply reformulate the sum to express it in terms of the symmetric state which, in the basis $\lla{\hat{B}^\dagger_{\jj_1}\ket{\mathrm{Mott}},\ldots, \hat{B}^\dagger_{\jj_{N_e}}\ket{\mathrm{Mott}}}$, writes as $\lambda_s=\pa{1\ldots 1}/\sqrt{N_e}$. Then, 
\begin{align}
N_e \Delta^2[y]&=\sum_{m} \abs{\sum_{\sss\in \lla{\jj}} \sum_{\rr \in \lla{\jj}; \rr\neq\sss}  \lambda_{m,\sss} x_{\sss\rr}}^2 = \sum_{m} \abs{\bkev{\mathds{1}}{X}{\lambda_m}}^2
=\norm{X\ket{\mathds{1}}}^2 - \abs{\bkev{\mathds{1}}{X}{\lambda_{s}}}^2 \\
&=\sum_{\sss\in \lla{\jj}} \pa{\sum_{\rr \in \lla{\jj}; \rr\neq\sss}  x_{\sss\rr}}^2 
-\frac{1}{N_e}\pa{\sum_{\sss\in \lla{\jj}} \sum_{\rr \in \lla{\jj}; \rr\neq\sss}  x_{\sss\rr}}^2 \,,
\end{align}
where $(X)_{\sss\rr}=x_{\sss\rr}$, and $\lambda_m=\pa{\lambda_{m,\jj_1}\ldots \lambda_{m,\jj_{N_e}}}$. The right hand side of the previous equation corresponds to $N_e$ times the variance of an homogeneous distribution of variables $y_\sss=\sum_{\rr \in \lla{\jj}; \rr\neq\sss}  x_{\sss\rr}$, with $\sss\in \lla{\jj}$.  It is therefore null when all fermions are equidistant, and the upper-bound is reached in the worst-case scenario of two fermions at distance 1 and the rest at infinite separation from each other. In this most-unfavourable situation, the latter expression reads as, $\pa{2-4/N_e}\pa{4\pi J_A}^{-2}\approx \co{2\pa{2\pi J_A}^{2}}^{-1}$. This contribution is therefore null for the two-fermion case. For many fermions it reduces the condition to $\varepsilon_{A,ii}=\left(\frac{V\st{III}}{\rho_M J_c }\right)^2 G(\{\jj\})\ll 1$ with,
\begin{equation}
    G(\{\jj\})=\pa{2\pi J_A N_e}^2\Delta^2[y]\ll N_e/2\,,
\end{equation}
as used in Eq.~\eqref{eq:wcond2}. The population of antisymmetric states in positions not occupied by the fermions is already bounded with these conditions.

\item One should also note that the fermionic hopping can also couple to symmetric to the antisymmetric states. This leads to an additional error that can be bounded by:
\begin{align}
    \varepsilon_f &=\sum_m \bar\varepsilon_f\pa{\ket{\phi^{(0)}_{\bot,m}}}\approx \sum_{m} \abs{t_F \frac{\braket{\phi^{(0)}_{\bot,m,\lla{\jj+1}}}{\phi^{(0)}_{+,\lla{\jj}}}}{\rho_M J_c}}^2\\
&=\pa{\frac{t_F}{\rho_M J_c}}^2\pa{1-\abs{\braket{\phi^{(0)}_{+,\lla{\jj+1}}}{\phi^{(0)}_{+,\lla{\jj}}}}^2}=\pa{\frac{t_F}{\rho_M J_c}}^2\pa{1-\frac{N_e-1+\mathcal{F}\st{III}}{N_e}}\approx \pa{\frac{t_F}{\rho_M J_c}}^2\frac{\text{a}}{L\st{III}N_e}
\end{align}
which corresponds to inequality~\eqref{eq:wcond4}. Note that we have assumed that the nearest neighbor in $\lla{\jj+1}$ is not occupied by a fermion, which is valid in the limit $\rho_M\ll 1$. In the last approximation, we have focused on the regime $L/\st{III}/\text{a}\gg 1$ where the Coulomb interaction dominates the Yukawa potential. This allows for a more relaxed condition than the one obtained when the effect of the Franck-Condon coefficient is neglected, as considered in \cite{arguello2019analogue}.
\end{itemize}

\section{Numerical methods for multi-electronic systems.}\label{ap:nummethods}

 To numerically capture both the geometry of the atom[molecule] and the interactions with other electrons in our analog simulator, we project the discretized fermionic Hamiltonian $\hat{H}_e$ in Eqs.~\eqref{subeq:kin}-\eqref{subeq:rep}, on a set of atomic[molecular] orbitals $\{\phi_i\}_{i=1}^{n}$ composed of two types of orbitals:
 \begin{itemize}
     \item Single-electron orbitals, corresponding to the $n$ first lowest energy eigestates of a single electron attracted to the same nuclear configuration. That means only the terms in Eqs.~\eqref{subeq:kin} and \eqref{subeq:nuc} in $\hat{H}_e$.
     \item Electronic orbitals that interact with an average-charge caused by the rest of electrons. For the case of two-electrons that we benchmark in this work, these Hartree-Fock-like orbitals are iteratively calculated by adding to the nuclear and kinetic terms in Eqs.~\eqref{subeq:kin} and \eqref{subeq:nuc} the repulsion due to the lowest-energy state obtained in the previous iteration given by Eq.~\eqref{subeq:rep}.
 \end{itemize}
 
Combining both sets of orbitals, the total basis is orthogonalized using Gram-Schmidt algorithm. The projected fermionic Hamiltonian then reads as,
	\begin{equation}
	\label{eq:proyHamiltonian}
	\hat H_e=\sum_{i,j,r,s=1}^{n} h_{ijrs}\ket{\phi_i\;\phi_j} \bra{\phi_r\; \phi_s}\,,
	\end{equation}
	where $h_{ijrs}=\bra{\phi_i\; \phi_j} \hat H_e\ket{\phi_r \; \phi_s}$, and $\ket{\phi_i\;\phi_j}$ denotes the product of the single-fermion states $\ket{\phi_i}\otimes \ket{\phi_j}$. A complete set would include $N$ independent orbitals in the basis, while we are computationally restricted to a dozen of orbitals when calculating the ground state. This limitation does not have any impact on the experiment, but it is desirable to estimate the imprecision made in this numerical benchmark. The success of this strategy then depends on how accurately the interactions in the Hamiltonian are captured by the orbitals included in this set and, therefore on the number and type of orbitals that we include in the truncated basis. In the following, we illustrate the application of this approach to the species benchmarked in the main text: atomic Helium, and molecular HeH$^+$.

	\subsection*{Numerical benchkmarking of atomic Helium}
	\label{ap:helium}
	In the case of Helium, one can explore the ortho-, and parahelium ground-states by restricting the projected Hamiltonian~\eqref{eq:proyHamiltonian} to the corresponding symmetry sectors. As para(ortho)helium is characterized by a(n) (anti)symmetric spin configuration, their spatial configuration needs to be antisymmetric(symmetric) due to their fermionic character. While this will be naturally ensured by the fermionic nature of our atomic simulator, the computation cost of the numerical calculation gets simplified by imposing these symmetries. In particular, one can define the reduced Hamiltonian 
	\begin{equation}
	\label{eq:redHamiltonian}
	\left. \hat H_e \right|_\text{para(ortho)}=\sum_{i,r}^{n}\sum_{j\geq i, s\geq r}^{n} h_{ijrs}^\text{para(ortho)}\ket{\ket{\phi_i\;\phi_j}}^\text{para(ortho)} \bra{\bra{\phi_r\; \phi_s}}^\text{para(ortho)} \,,
	\end{equation}
	where $\ket{\ket{\phi_i\;\phi_j}}^\text{para(ortho)}=\co{\ket{\phi_i\;\phi_j}+(-)\ket{\phi_j\;\phi_i}}/\sqrt{2}$, and $h_{ijrs}^\text{para(ortho)}=h_{ijrs}+(-)h_{jirs}$, where we have used the identity, $h_{ijrs}=h_{jisr}$.

We should emphasize that this projection on a single-particle basis is just a numerical strategy that enables us to numerically benchmark the model, but does not have any impact on the experimental implementation of the proposed analog simulator.

\begin{figure}[tbp]
	\centering
	\includegraphics[width=0.5\linewidth]{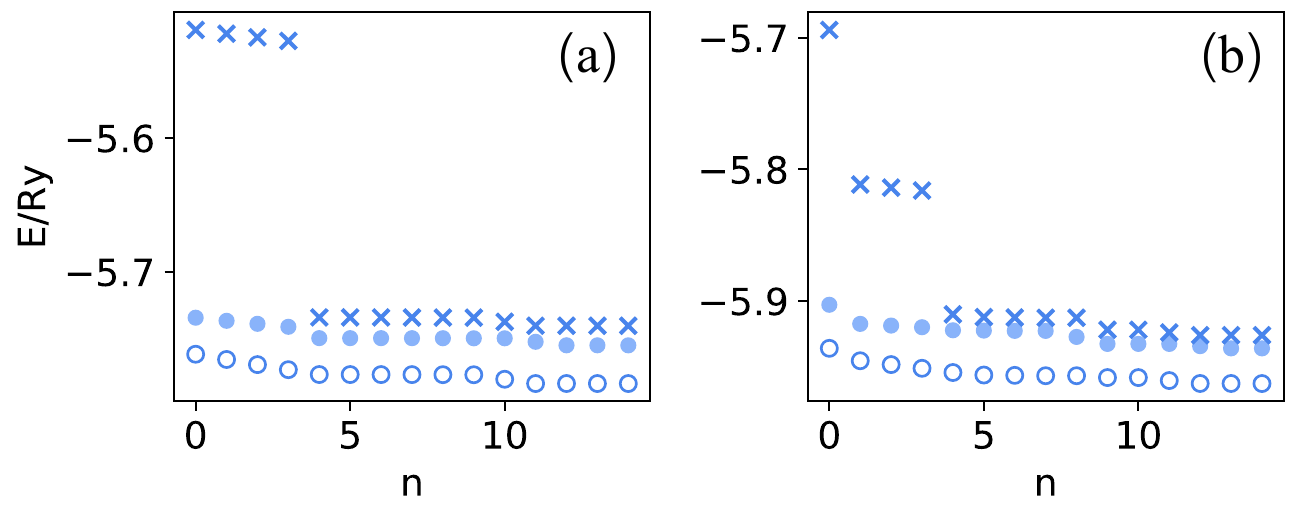}
	\caption{(a) Numerical benchmarking for the ground-state energy of He for an increasing number of orbitals in the projective basis for the example of tunneling rate, $t_f/V_0=6$ and $N=75^3$. In particular, we compare the lowest energy of the projected Hamiltonian for basis composed of the $n$ lowest-energy single-electron orbitals of He$^+$ (crossed markers), Hartree-Fock orbitals constructed as described in this section (coloured round markers), or a combination of the n-th first of them (contoured round markers). (b) We repeat this analysis for the lowest-energy of HeH$^+$, for an internuclear distance $d/a_0=1.5$ simulated with a separation of $d/\text{a}=15$ sites, and $N=75^3$.}
	\label{fig:conver}
\end{figure}

In Fig.~\ref{fig:conver}(a), we analyze the convergence of result by calculating plot the lowest energy of He atoms as a function of the type and number of orbitals included in the basis. As expected, orbitals obtained using the HF approach (coloured round markers) diminish more easily the ground-state energy than single electron orbitals (crossed markers). A combination of both basis (contoured round markers) show the greatest reduction. For the convergence of the results shown along the text, we have chosen 30 orbitals: 15 coming from the single electron calculation, and 15 obtained with the described HF method, which show energy variations smaller than the energy error provided.

\subsection*{\texorpdfstring{Numerical benchkmarking of molecular HHe$^+$}{Numerical benchkmarking of molecular HHe+}}

In this case, the chosen Bohr-radius $\text{a}_0/\text{a}$ modifies the effective internuclear separation $d/\text{a}_0$. To explore the effect of discretization, for a given physical distance $d/\text{a}_0$, we then modify the nuclear separation $d/\text{a}$ taking integer values, and adjust the effective Bohr-radius $\text{a}_0/\text{a}$ accordingly. In Fig.~\ref{fig:molHeH}, this process is repeated for lattice sizes $N=75^3$ and $N=100^3$, and the extrapolation method is then used to extract the best estimation of the ground-state energy in the continuum from our Hamiltonian in the lattice for each value of $d/\text{a}_0$. 

As it occurred in the case of He, a single-electron base obtained from a Hartree-Fock approach is used to solve the discretized molecular Hamiltonian in a a projected basis. In Fig.~\eqref{fig:conver}(b) we benchmark its convergence with the number of orbitals, observing energy variations smaller than the energy error for the choice of 15 orbitals using the single electron calculation, and 15 obtained with the described HF method.
\end{document}